\documentclass[fleqn,usenatbib]{mnras}

\usepackage{newtxtext,newtxmath}

\usepackage[T1]{fontenc}
\usepackage{ae,aecompl}

\usepackage{graphicx}	
\usepackage{amsmath}	
\usepackage{color, colortbl}
\usepackage{caption}

\usepackage{soul} 
\usepackage{xargs} 
\usepackage[pdftex,dvipsnames]{xcolor}  
\usepackage{amsmath}
\usepackage{longtable}
\definecolor{shade}{gray}{0.9}

\newcommand{\astar }{$A_{\star}$}

\title[Long GRB environments]{Towards an understanding of long gamma-ray burst environments through circumstellar medium population synthesis predictions}

\author[A. A. Chrimes et al.]{A. A. Chrimes,$^{1}$\thanks{E-mail: a.chrimes@astro.ru.nl}
B. P. Gompertz,$^{2,3}$
D. A. Kann,$^{4}$ 
A. J. van Marle,$^{5,6}$
J. J. Eldridge,$^{7}$
P. J. Groot,$^{1,8,9}$ \newauthor 
T. Laskar,$^{1}$ 
A. J. Levan,$^{1,3}$
M. Nicholl,$^{2}$
E. R. Stanway,$^{3}$
and K. Wiersema$^{10}$ 
\\
$^{1}$Department of Astrophysics/IMAPP, Radboud University Nijmegen, P.O. Box 9010, 6500 GL Nijmegen, The Netherlands\\
$^{2}$Institute of Gravitational Wave Astornomy and School of Physics and Astronomy, University of Birmingham, Birmingham, B15 2TT, UK \\
$^{3}$Department of Physics,  University of Warwick, Gibbet Hill Road, Coventry, CV4 7AL, UK\\
$^{4}$Instituto de Astrofísica de Andalucía (IAA-CSIC), Glorieta de la Astronomia s/n, 18008 Granada, Spain \\
$^{5}$Laboratoire Univers et Particules de Montpellier (LUPM), Universit\'e de Montpellier \& CNRS, Place Eug\'ene
Bataillon, 34095 Montpellier Cedex 05, France\\
$^{6}$ ELI Beamlines, Institute of Physics, Academy of Sciences,
25241 Dolní Břežany, Czech Republic\\
$^{7}$Department of Physics, University of Auckland, Private Bag 92019, Auckland, New Zealand\\
$^{8}$Inter-University Institute for Data Intensive Astronomy, Department of Astronomy, University of Cape Town, Private Bag X3, Rondebosch 7701, South Africa \\
$^{9}$South African Astronomical Observatory, P.O. Box 9, 7935 Observatory, South Africa \\
$^{10}$Physics Department, Lancaster University, Lancaster, LA1 4YB, UK \\
}

\date{Accepted XXX. Received YYY; in original form ZZZ}

\pubyear{2022}

\begin{document}
\label{firstpage}
\pagerange{\pageref{firstpage}--\pageref{lastpage}}
\maketitle

\begin{abstract}
The temporal and spectral evolution of gamma-ray burst (GRB) afterglows can be used to infer the density and density profile of the medium through which the shock is propagating. In long-duration (core-collapse) GRBs, the circumstellar medium (CSM) is expected to resemble a wind-blown bubble, with a termination shock separating the stellar wind and the interstellar medium (ISM). A long standing problem is that flat density profiles, indicative of the ISM, are often found at lower radii than expected for a massive star progenitor. Furthermore, the presence of both wind-like environments at high radii and ISM-like environments at low radii remains a mystery. In this paper, we perform a `CSM population synthesis' with long GRB progenitor stellar evolution models. Analytic results for the evolution of wind blown bubbles are adjusted through comparison with a grid of 2D hydrodynamical simulations. Predictions for the emission radii, ratio of ISM to wind-like environments, wind and ISM densities are compared with the largest sample of afterglow-derived parameters yet compiled, which we make available for the community. We find that high ISM densities of $n \sim 1000$\,cm$^{-3}$ best reproduce observations. If long GRBs instead occur in typical ISM densities of $n \sim1$\,cm$^{-3}$, then the discrepancy between theory and observations is shown to persist at a population level. We discuss possible explanations for the origin of variety in long GRB afterglows, and for the overall trend of CSM modelling to over-predict the termination shock radius.
\end{abstract}

\begin{keywords}
${\gamma}$-ray burst: general -- stars: winds, outflows -- stars: Wolf–Rayet
\end{keywords}



\section{Introduction}
Gamma-ray bursts (GRBs) are among the most energetic explosions in the Universe. Long-duration GRBs (loosely defined as having 90\% of their prompt ${\gamma}$-ray flux arrive over more than 2s) predominantly arise from core-collapse events with a bias towards low metallicity, as evidenced by their host galaxy environments and associations with stripped-envelope type Ic supernovae \citep[][]{1998Natur.395..670G,2006Natur.441..463F,2009ApJ...691..182S,2016SSRv..202...33L,Cano2017AdAst,2019EPJA...55..132F}. The canonical long-duration collapsar model invokes jets that are launched by a rapidly spinning compact object, born in a core-collapse supernova \citep[][]{1993ApJ...405..273W}. Strong shocks in the jet produce ${\gamma}$-rays and X-rays through synchrotron radiation \citep[and TeV emission through other radiative processes,][]{2019Natur.575..455M,Magic2019Nature2}. The jets are initially highly relativistic and strongly beamed, but as they collide with the immediate circumstellar medium (CSM), they decelerate and expand \citep{1976PhFl...19.1130B,2001ApJ...562L..55F}. The spectral peak moves from the X-rays through to the radio bands, while the light curve chromatically fades, and a quasi-achromatic jet break occurs when the beaming cone of the jet approaches $1/ \Gamma \approx \theta_{\rm jet}$, the jet opening angle \citep[although this jet break can occur so long post-burst that the afterglow has faded beyond detectability limits,][]{1999ApJ...525..737R}.

The temporal and spectral behaviour of the afterglow prior to the jet break can be used to infer the density\footnote{Unless otherwise stated, by density we refer to the mass density divided by the proton mass, i.e. the approximate number density of particles in the medium.} profile of the medium through which it is propagating \citep{1998ApJ...497L..17S,2000ApJ...536..195C,2000ApJ...543...66P,2002ApJ...571..779P}. Afterglow studies have found that long GRBs typically occur in either `wind-like' environments, with an r$^{-2}$ density profile as expected for stellar winds close to the progenitor, or constant-density environments, often attributed to the interstellar medium further out \citep[e.g.,][]{2000ApJ...543...66P,2001ApJ...559..123H,2001ApJ...554..667P,2002ApJ...571..779P,2002ApJ...577..155Y,2004ApJ...606..369C,Chandra08,2010ApJ...711..641C,Cenko11,Schulze2011AA,Laskar14,2018ApJ...866..162G,2020arXiv200906740S,2021arXiv210607169S}. For example, lightcurves decline more rapidly in wind-like media pre-jet break, in the spectral regime $\nu_{m} < \nu < \nu_{c}$ (between the peak and cooling synchrotron break frequencies). The transition from a wind-like to constant-density regime occurs at $r_\mathrm{wind}$, the termination shock radius of the stellar wind. Along with $r_\mathrm{wind}$, the quantity \astar\ , which parameterises the density of the stellar wind, can be determined for wind-like bursts, while the constant density $n$ can be inferred for interstellar medium (ISM)-like fits. ISM-like environments close to the progenitor are found in short GRBs \citep[e.g.,][]{2020MNRAS.495.4782O}. However, this is expected in short GRBs because the binary neutron star progenitors are not shedding mass shortly before the GRB. Furthermore, they have likely received natal kicks, sending them into the wider ISM or even intergalactic medium \citep[IGM,][]{2010ApJ...722.1946B,2013ApJ...776...18F,2014MNRAS.437.1495T}.

In long GRB afterglows, it is not always the case that one environment type is clearly preferred from model fitting \citep[e.g.][]{2018ApJ...866..162G}. Furthermore, where ISM or wind fits are preferred, the observed ratio of these populations varies from study to study. \citet{Schulze2011AA} find an ISM to wind ratio (among {\em Swift}-detected long GRBs) of 3, indicating that in the majority of events, $r_\mathrm{wind}$ lies close-in, between $10^{-3}$--1\,pc, with a small and possibly distinct population of events where it lies further out on average ($>10^{-1}$\,pc). \citet{Schulze2011AA} also find that $r_\mathrm{wind}$ in wind-like fits lies further from the progenitor than in ISM fits, consistent with the wind bubble model. Conversely, \citet{2018ApJ...866..162G} find a near equal ratio of ISM and wind-like environments among {\em Fermi} long GRBs, and that the bursts with the highest emission radii at a rest-frame time of 11 hours are in wind-like environments. This is hard to explain without there being substantial variety in the wind properties or ISM environments of the progenitors.

Ever since the core-collapse of massive, low-metallicity stars became the favoured model for long GRBs, there have been efforts to use stellar evolution models that satisfy these criteria to determine if their mass loss rates and wind speeds could reproduce the CSM properties observed in GRB afterglows \citep{2006MNRAS.367..186E,2006A&A...460..105V,2008A&A...478..769V}. A problem arises, however, because the emission radius of many long GRBs at the time of observation is lower than the predicted termination shock radii from both analytic approximations and hydrodynamical simulations \citep[$\sim1$ pc is expected for Wolf-Rayet stars in typical galactic ISM densities of $\sim$1\,cm$^{-3}$, e.g.][]{2006MNRAS.367..186E,2006A&A...460..105V,2007MNRAS.377L..29E,2007MNRAS.376L..57K,2008A&A...478..769V,Schulze2011AA,2018ApJ...866..162G}. At low distances from the star, the density profile is expected to be wind-like, with the radius of the termination shock dependent on a pressure balance between the ISM  and the wind. This is set by the mass loss rate history of the progenitor, the wind speeds, the ISM density, the metallicity and temperature of the gas. However, at radii expected to be wind-like for massive star progenitors in a typical galactic ISM, many afterglows exhibit evolution consistent with a constant density medium. Efforts to model wind-blown bubbles, both with analytic solutions and hydrodynamical simulations, have struggled to achieve the low $r_\mathrm{wind}$ values of 0.1-1\,pc inferred from some long GRB afterglows. Wolf-Rayet progenitors, for example, should drive strong winds \citep[which are seen in GRB afterglow fits, e.g.][]{2021arXiv210614921A}, pushing $r_\mathrm{wind}$ further out and favouring wind-like media at such small distances from the star.

\citet{2008ApJ...685..344P} provide strong constraints on the environments of six long duration bursts. They determine from the velocity structure of ionised nitrogen (N {\sc V}) absorption lines that the afterglow had reached an ISM-like region within 10\,pc in every case. This argument follows from N {\sc V} only being expected out to a certain radius (dependent on the ionising flux of the GRB), and narrow lines being expected in the ISM, not in the high-velocity outflows in the wind region. Therefore, if N {\sc V} is expected at $r<10$\,pc only, but the lines are narrow, this implies that the ISM has been reached within $10$\,pc. Seemingly in contradiction, broad absorption features ($>100$\,km\,s$^{-1}$, indicating a wind-like environment) have been seen out to $\sim$100\,pc in other GRB afterglows \citep[][and references therein]{2012A&A...547A..83G}. A possible solution to this conflict is the stalled wind (also referred to as the shocked wind) region. In this picture, the N {\sc V} lines originate from the stalled wind (at or inside 10\,pc), and beyond the inner wind shock  (driven by a Wolf-Rayet or He star) lies the high velocity outflow from a previous supergiant or luminous blue variable (LBV) phase. Electron densities (rather than particle densities more generally) of a few hundred cm$^{-3}$ have also been found at the locations of GRBs via spatially-resolved electron density-sensitive emission line measurements \citep[e.g.][]{2004ApJ...611..200P,2007A&A...464..529W,2014A&A...562A..70M}. However, these measurements are biased towards regions of high flux and the sensitivity to lower densities declines rapidly.

Later, \citet{2018ApJ...866..162G} calculated \astar\ , $n$ and the emission radius $r_{\rm emit}$ for a sample of 56 bursts that were observed with the {\it Fermi} satellite. {\it Fermi}-GBM  \citep{2009ApJ...702..791M} is sensitive to higher energy photons than the {\it Neil Gehrels Swift Observatory's} (\emph{Swift} hereafter, \citealt{Gehrels04}) Burst Alert Telescope \citep[BAT,][]{2005SSRv..120..143B}, and so more frequently captures the GRB peak photon energy E$_{\rm p}$, which allows for better estimation of the total prompt energy and hence the emission radius \citep{2005ApJ...633.1018P}. \citet{2018ApJ...866..162G} found that $r_\mathrm{wind}$ likely spans the range $0.1-10$ pc, based on the emission radii at 11 hours (rest frame).

No correlation between emission radius and environment type was found; in some cases, constant-density media are found close to the star, in others, wind-like profiles persist to ${\sim}$10\,pc. This suggests that there is a wide range of progenitor wind properties, environmental densities, or both. If not, then the typical observation time of ${\sim}$11 hours must be (coincidentally) close to the mean termination shock crossing time, in order to produce the 50:50 split between ISM and wind seen in the sample. In this case, the signature of the jet crossing the shock - typically expected to be a shallowing temporal index, possibly accompanied by flares - should be frequently seen \citep{Dai02,2002A&A...396L...5L,2006MNRAS.367..186E,2006ApJ...651.1005S}. While there have been hints of such transitional signatures, they remain rare and ambiguous \citep{2003ApJ...591L..21D,2009MNRAS.400.1829J,2011RAA....11.1046F,2011A&A...531A..39N,2015ApJ...810...31V,2020arXiv200802445L,2021arXiv210414080L}, although it may also be the case that flares are not produced, even when the blast wave encounters an abrupt density increase \citep{2013ApJ...773....2G}.

Since these questions were first explored, stellar evolution and hydrodynamic wind modelling have advanced, and over a decade more of {\it Fermi}, {\it Swift} and follow-up observations have been obtained. In this paper, we employ BPASS stellar evolution models \citep{2017PASA...34...58E,2018MNRAS.479...75S} for GRB progenitors to perform a `circumstellar medium population synthesis'. The models have been selected for their agreement with long GRB rates, host galaxy metallicities, hydrogen and helium poor chemical compositions and theoretical expectations for their pre-collapse rotation \citep{2020MNRAS.491.3479C}. We calculate \astar\ and $r_\mathrm{wind}$ for each model in a range of ISM densities using analytic approximations, correcting for and determining the uncertainty in these results by comparing to results obtained from a smaller grid of hydrodynamical simulations. We test these predictions against a large multi-wavelength afterglow dataset, comprising 75 long-duration bursts. We ask whether the observed diversity and distributions of \astar\ , $n$ and $r_{\rm emit}$ can be reproduced by CSM population synthesis, and if so, under which ISM conditions.

The paper is structured as follows. In Section \ref{sec:obsv} we describe the afterglow dataset, data reduction and parameter inference. Section \ref{sec:models} gives an overview of the GRB progenitor models used. In Section \ref{sec:analytic} we describe how we model stellar winds and stellar wind bubbles, before performing a wind bubble population synthesis with analytic solutions. In section \ref{sec:hydro} we quantify the effect of assumptions in this analytic model by running a grid of hydrodynamical simulations and examining trends in the differences between the analytic and numerical approaches. Section \ref{sec:results} compares our corrected distributions of \astar, $r_\mathrm{wind}$, and $n$ to the observational sample. We discuss these results in Section \ref{sec:discuss} and present conclusions in Section \ref{sec:conclusion}. Magnitudes are quoted throughout in the AB system \citep{1983ApJ...266..713O}, and a $\Lambda$CDM cosmology with $h$ = 0.696, ${\Omega}_\mathrm{M}$ = 0.286 and ${\Omega}_{\Lambda}$ = 0.714 is assumed \citep{2014ApJ...794..135B}.

\section{Afterglow Sample}\label{sec:obsv}
\subsection{Sample construction}
Our sample of long GRB afterglows is compiled following the criteria of \cite{Kann2006Apj}. We require that our selected GRBs have a known redshift and multi-epoch afterglow detections, which typically ensures that they have at least optical and X-ray data (although not always, a few cases, even in the \emph{Swift} era, have optical afterglows only). X-ray data are taken by \emph{Swift's} X-ray Telescope \citep[XRT; ][]{Burrows05}. Light curve data in the $0.3$ -- $10$\,keV energy range are downloaded from the UK \emph{Swift} Science Data Centre \citep[UKSSDC;][]{Evans07,Evans09} and converted to 1\,keV flux density \citep[cf.][]{Gehrels08}. Optical data are collected from the samples presented in \citet[][2022a,b, in prep.]{Kann2006Apj,Kann2010Apj,Kann2011Apj}. For almost all cases, multi-band UV/optical/NIR fitting of the afterglow light curves, combined with the known redshift, allows determination of the dust along the line-of-sight and therefore a correction for intrinsic extinction. In case of achromatic evolution, the observations are synthesised into composite $r$-band light curves to maximise data density and temporal coverage. These light curves are furthermore, where applicable, corrected for the host galaxy and supernova contributions, leaving pure afterglow light curves. For full details of the sample compilation, we direct the reader to \citet[][2022a,b, in prep.]{Kann2006Apj,Kann2010Apj,Kann2011Apj}. The requirement for optical afterglow detections biases the sample against dark bursts, which have lower than expected optical emission based on extrapolation of the X-ray spectral energy distribution \citep{1998ApJ...493L..27G,2004ApJ...617L..21J,2009ApJ...699.1087V}. Typically, such bursts are in dusty environments, with a small proportion being dark due to rest-frame UV absorption by neutral hydrogen at high redshift \citep[e.g.][]{2011A&A...526A..30G}. The likely effect of their exclusion is discussed later. In addition to the optical and X-ray data described above, radio data are collected from the literature. See Table~\ref{app:fits} for cited works.

\subsection{Afterglow parameter fitting}
Afterglow light curves are fit with analytic approximations of evolving synchrotron spectra appropriate for relativistic jets decelerating in the ambient environment and forming shocks \citep{1998ApJ...497L..17S,2000ApJ...536..195C,2000ApJ...543...66P,2002ApJ...571..779P,Granot02}. The scintillation of radio photons by electrons in the Milky Way is accounted for in our fits by adding the expected amplitude of scintillation in quadrature with the observational errors. The amplitude of scintillation is calculated using the method of \citet{Gompertz17}, which is based on the model of \citet{Goodman97} and uses the free electron distribution model of \citet{Cordes02}. Our model parameterises the afterglow in terms of its isotropic equivalent kinetic energy ($E_{\rm k, iso}$), the underlying distribution of electron energies, which is assumed to be a power-law with an index $p$, the fraction of energy that is contained in the radiating electrons ($\epsilon_e$), the fraction of energy that is contained in the magnetic fields ($\epsilon_B$), the density of the circumburst medium ($n$ for an ISM-like $\rho \propto r^0$ or $A_*$ for a wind-like $\rho \propto r^{-2}$ environment) and the half-opening angle of the jet $\theta_o$. The value of $\epsilon_e$ is found to have a very narrow distribution \citep{Beniamini17}, and we fix it to $\epsilon_e = 0.1$ to reduce the number of free parameters. Our model therefore has 5 free parameters: $E_{\rm k,iso}$, $p$, $\epsilon_B$, $n$/$A_*$ and $\theta_o$. We do not account for the possibility of density profiles intermediate between the theoretically expected ISM and wind cases due to the already large parameter space and sample size \citep[see however][]{Starling08}.

\begin{table}
    \centering
    \begin{tabular}{ccccc}
    \hline\hline
        Parameter & Initial & Bounds & Units & Prior \\
    \hline
        $E_{\rm k,iso}$ & $10^{52}$ & $10^{49}$ -- $10^{56}$ & erg & Log flat \\
        $p$ & $2.2$ & $2.0$ -- $3.0$ & & Flat \\
        $\epsilon_B$ & $10^{-2}$ & $10^{-5}$ -- $10^{-0.5}$ & & Log flat\\
        $n$ & $1$ & $10^{-5}$ -- $10^2$ & cm$^{-3}$ & Log flat \\
        $A_*$ & $1$ & $10^{-5}$ -- $10^2$ & $5\times10^{11}$\,g\,cm$^{-1}$ & Log flat \\
        $\theta_o$ & $0.2$ & $0.02$ -- $0.5$ & rad & Flat \\
    \hline\hline
    \end{tabular}
    \caption{Priors for our five free parameters ($n$ and $A_*$ are interchanged depending on environment type). 100 walkers were initialised either evenly spaced within the respective bounds or clustered around the initial value.}
    \label{tab:priors}
\end{table}

Model fits are performed using the Markov Chain Monte Carlo (MCMC) method with the Python package {\sc emcee} \citep{emcee}. We utilise 100 walkers and 20,000 steps, discarding the first 5,000 as burn-in. We employ a cut of $t > 0.2$ days on the data in order to avoid the rapid variability often seen in the early evolution of GRB afterglows, which is likely due to prolonged engine activity and is not accounted for in our models. This cut also mitigates the influence of reverse shocks which may be present in radio observations at early times and are not accounted for in our model. We perform four runs for each GRB, two per environment type, one with walkers initially clustered around a fixed starting point and one where they are evenly spaced within the allowed parameter limits (Table~\ref{tab:priors}).

Convergence is assessed via inspection of the light curves and corner plots. \citet{2018ApJ...866..162G} demonstrated that even in cases with good data coverage, analytic approximations of GRB afterglows are not always able to converge on consistent underlying physics and require a reasonably large `ignorance' parameter. In the present work, we are interested in the broad distribution of GRB environment densities, and hence we require that our models match the broad evolution of the afterglows, but do not penalise them for failing to reproduce smaller-scale variations that may be unobtainable with simplified theoretical assumptions, or due to processes not included in the model (e.g. energy injection, `clumpy' density profiles, etc). The designation of wind or ISM is made by inspecting the relative affinity of the models to the available observations, including whether the preferred solution is able to point to a clear maximum in the density posterior and other fit parameters.

Fits were performed on {\sc gotohead}, the computing cluster of the Gravitational-wave Optical Transient Observer (GOTO) collaboration \citep{Dyer20,Gompertz20,Steeghs22}. The full table of results is shown in Table~\ref{app:fits}, with some example fits and corner plots in Figures~\ref{app:081007} and \ref{app:071003}. Fits and corner plots for every burst in the sample are available as supplementary material on the journal website. While the results provide representative sampling of the distributions of GRB afterglow parameters, we caution the reader against over-interpreting individual fits, which are necessarily the product of simplified physical models applied to complex systems (a general problem when fitting models to GRB afterglows). From a sample size of 74, we find a ratio of ISM to wind-like environments of 45/29 = 1.55$\pm$0.37 (Poisson uncertainties). Previous estimates for this ratio are in the range $\sim$1--3, with the precise value likely dependent on sample selection effects and the modelling methods used \citep{Schulze2011AA,2018ApJ...866..162G,2022ApJ...927...84Z}.

\section{GRB Progenitor Models}\label{sec:models}
To model the CSM of long GRB progenitors, we first need stellar evolution models, with properties such as the mass-loss rate as a function of time. The binary stellar evolution models we use are selected according to the criteria of \citet{2020MNRAS.491.3479C} from BPASS \citep[Binary Population and Spectral Synthesis v2.2.1,][]{2017PASA...34...58E,2018MNRAS.479...75S}. The BPASS models explicitly include binary interactions such as mass transfer, and cover the full range of binary parameter space. The models are weighted in metallicity bins, according to their predicted occurrence in a 10$^{6}$ M$_{\odot}$ population \citep[based on observations of local stellar populations,][]{2017ApJS..230...15M}. In \citet{2020MNRAS.491.3479C}, tidal interactions are added to the models in post-processing, and the stellar rotations are tracked.  

Two pathways are assumed to produce viable long GRB progenitors. The first is quasi-homogeneous evolution (QHE), where a secondary star ($>20$M$_{\odot}$) is spun up by accretion, undergoing increased rotational mixing. This more efficiently burns the hydrogen and helium, and moves the star blue-wards on the Hertzsprung-Russel diagram. If this occurs at low metallicity ($<0.004$ by mass fraction), we assume that the star maintains a high enough spin all the way to core-collapse to be a long GRB progenitor \citep[as line-driven winds and therefore angular momentum loss are low,][]{2006A&A...460..199Y}.

Alternatively, because GRBs are known to occur in hosts with metallicities higher than 0.2$Z_{\odot}$ \citep[e.g.][]{2009AIPC.1133..269G,2010AJ....140.1557L,Elliott2013AA,2015A&A...579A.126S,2018A&A...616A.169M}, the possibility of tidally spun-up progenitors in binaries is considered \citep[see e.g.][for further discussion of the predictions and implications of this progenitor channel]{2004ApJ...607L..17P,2004MNRAS.348.1215I,2005A&A...435..247P,2008A&A...484..831D,2022A&A...657L...8B,2022arXiv220405013A}. In these cases, tidal interactions with a companion can maintain rotation where mass loss would otherwise spin the star down. The tidal models are selected if,
\begin{enumerate}
    \item The chemical composition at core-collapse is reflective of type Ic supernovae, i.e. low hydrogen and helium atmospheric abundances,
    \item A black hole is produced in the collapse,
    \item The surface angular momentum of the star is above a threshold, tuned to reproduce the observed long GRB rate. 
\end{enumerate}
The two pathways (QHE and tides) were together able to reproduce both the observed GRB rate \citep[using the metallicity-dependent star formation history of][]{2006ApJ...638L..63L}, and independently, the metallicity distributions of the hosts \citep[e.g.][]{2019A&A...623A..26P}. Assuming rigid rotation (strong core-envelope coupling), the chosen angular momentum threshold also corresponds to specific angular momenta outside the core of $\sim 10^{16}$cm$^{2}$s$^{-1}$, close to the value required by collapsar accretion disc theory \citep{1993ApJ...405..273W,2005A&A...435..247P}.

We use the tide-adjusted models calculated in \citet{2020MNRAS.491.3479C}. These include `jumps' between models where tidal interactions modify the orbital period enough that the system orbital parameters are now better represented by a different model at that timestep. However, these model jumps are the exception rather than the norm, occurring in only $\sim$10 per cent of cases, and jumps are typically only to adjacent models in the model grid. Therefore, the mass loss rates as a function of time in the models used are similar (in most cases identical) to the standard BPASS models. Across the redshift range where afterglows can be studied in detail, the tidal pathway (metallicities up to $\sim Z_{\odot}$) contributes a factor of a few more to the total rate than the QHE channel ($Z<0.2 Z_{\odot}$). For further details, we refer the reader to \citet{2020MNRAS.491.3479C}, and to \citet{2019MNRAS.482..870E}, \citet{2020MNRAS.493L...6T} and \citet{2021arXiv211108124B} for further discussions of transient rate calculations with BPASS.

\section{Wind Modelling and Bubble Population Synthesis}\label{sec:analytic}
In this section we calculate, for each progenitor model, the wind termination shock radius and the wind density parameter \astar\ at a range of ISM densities. This CSM population synthesis is initially performed by application of analytic solutions to the BPASS models outputs. These results are then adjusted in Section \ref{sec:hydro} based on comparison to a smaller grid of hydrodynamical simulations. This approach avoids the computational cost of running hundreds of detailed simulations, while still benefiting from the improved accuracy they provide. We assume spherical symmetry throughout.

\begin{figure*}
\centering
\includegraphics[width=0.99\textwidth]{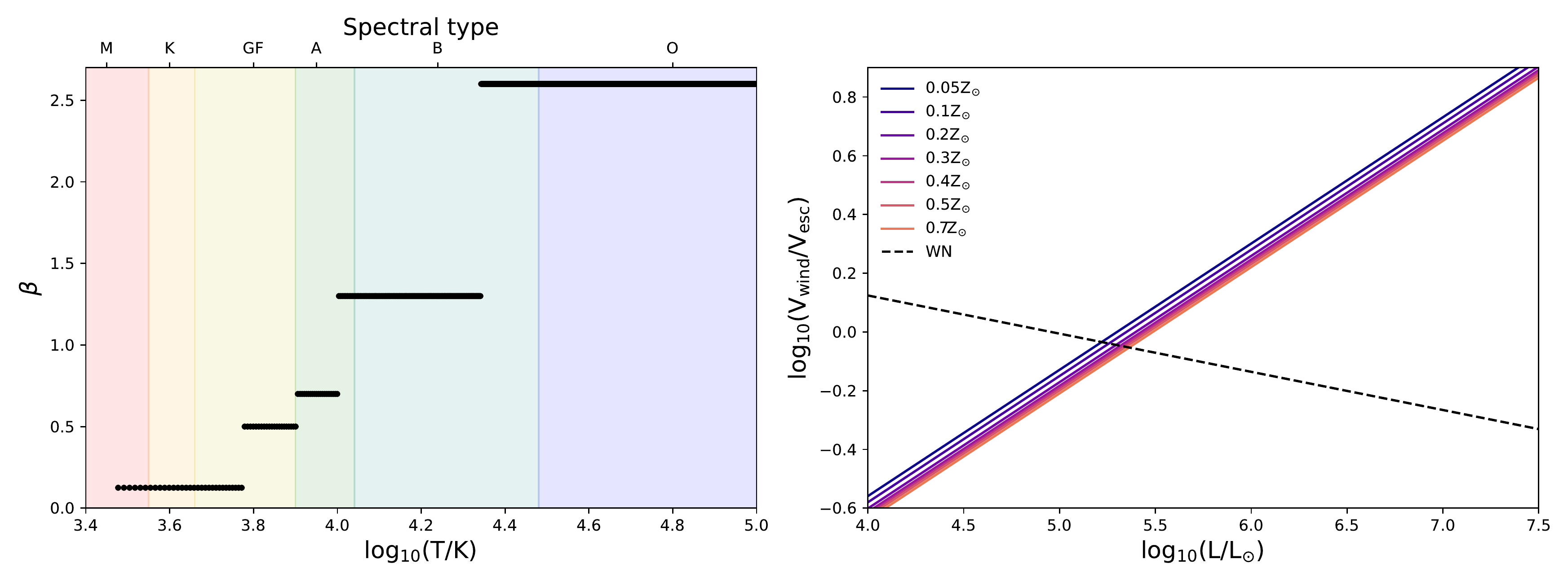}
\caption{Left: wind speeds in the optically thin regime, where ${\beta}$ is as defined in Equation \ref{eq:one}. We use the wind speeds of \citet{2006MNRAS.367..186E}, which are based on results from \citet{2002MNRAS.329..897H}, \citet{2001A&A...369..574V} and \citet{1995ApJ...455..269L}. The speeds calculated from the values shown here are modified by a factor ($Z/Z_{\odot}$)$^{0.13}$. The BPASS spectral type definitions are also indicated \citep{2017PASA...34...58E}. Right: optically thick winds (i.e. the Wolf-Rayet regime). We adopt the WC prescription of \citet{2000A&A...360..227N} for all Wolf-Rayet stars, as this best approximates (out of the WC and WN prescriptions) the results of detailed atmospheric modelling for all Wolf-Rayets \citep[including WC, WO, WN and WNh types,][]{2020MNRAS.499..873S}, except at the lower luminosity limit of Wolf-Rayet behaviour. The WC winds are shown for a range of metallicities, as is the WN wind (not used) at 0.5\,Z$_{\odot}$. }
\label{fig:Windspeeds}
\end{figure*}

\subsection{Stellar winds}
The BPASS model outputs include every quantity required to calculate wind and wind bubble parameters, except wind velocities. We adopt wind speeds which are similar to those used in the STARS models \citep[][on which the BPASS models are built]{2006MNRAS.367..186E}, with some alterations following the results of detailed helium star atmospheric modelling \citep{2020MNRAS.499..873S}. The BPASS mass loss rates themselves are not altered, as these are part of the stellar evolution models and are included with the model outputs.

\subsubsection{Main sequence and pre-Wolf-Rayet}
For stars that are not Wolf-Rayets, including those on the main sequence and all phases post-main sequence up to becoming a Wolf-Rayet, we implement a modified version of the wind speeds used by \citet{1995ApJ...455..269L}, \citet{2002MNRAS.329..897H}, \citet{2001A&A...369..574V} and \citet{2006MNRAS.367..186E}. The wind speed in the non-Wolf-Rayet regime is given by,
\begin{equation}\label{eq:one}
    V_\mathrm{w} = \sqrt{\beta V_\mathrm{esc}^{2}} \times \Big(\frac{Z}{Z_\odot}\Big)^{0.13}
\end{equation}
where $Z$ is the metallicity (by mass fraction), and $\beta$ parameterises the wind strength for different effective temperatures. We use the $Z^{0.13}$ metallicity modification of \citet{1992ApJ...401..596L} and \citet{2001A&A...369..574V}. The ${\beta}$ values used are shown in the left panel of Figure \ref{fig:Windspeeds}.

The escape velocity at the stellar surface is modelled in the following way. First, a cross-section is defined which depends on the hydrogen and helium atmospheric mass fractions,
\begin{equation}
    \sigma_\mathrm{e} = 0.401\Big(X+\frac{Y}{2}+\frac{Y}{4}\Big)\,\mathrm{cm}^{2},
\end{equation}
which appears in the Eddington factor, along with the mass and luminosity. This accounts for a reduction in {\it effective} escape velocity for stars close to the Eddington luminosity L$_\mathrm{Edd}$ (i.e., a particle in the wind gets a boost from radiation pressure). The Eddington factor is defined as,
\begin{equation}
    \Gamma = \frac{L_\star}{L_\mathrm{Edd}} = 7.66\times10^{-5}\sigma_\mathrm{e}\Big(\frac{L_\star}{M_\star}\Big),
\end{equation}
and the effective escape velocity is,
\begin{equation}
    V_\mathrm{esc} = \sqrt{\frac{2GM_\star}{R_\star}(1-\Gamma)},
\end{equation}
where M$_{\star}$ and R$_{\star}$ are the stellar mass and radius respectively.

\subsubsection{Wolf-Rayet winds}
For Wolf-Rayet stars, we employ the wind velocities of \citet{2000A&A...360..227N}, on which the BPASS v2.2.1 mass loss rates are also based. We identify Wolf-Rayet stars using the standard BPASS definitions \citep[based on temperature and surface abundances,][]{2008MNRAS.384.1109E,2017PASA...34...58E}. For WN stars (Wolf-Rayets with nitrogen-rich composition), the wind speed is given by,
\begin{equation}
    \mathrm{log}_{10}\Big(\frac{V_\mathrm{wind}}{V_\mathrm{esc}}\Big) = 0.61-0.13\mathrm{log}_{10}\Big(\frac{L_{\star}}{L_{\odot}}\Big)+0.30\mathrm{log}_{10}(Y)
\end{equation}
where Y is the helium mass fraction. We also have,
\begin{equation}
    \mathrm{log}_{10}\Big(\frac{V_\mathrm{wind}}{V_\mathrm{esc}}\Big) = -2.37+0.43\mathrm{log}_{10}\Big(\frac{L_{\star}}{L_{\odot}}\Big)-0.07\mathrm{log}_{10}(Z)
\end{equation}
for WC stars (Wolf-Rayets with carbon-rich composition).

These relations are shown in Figure \ref{fig:Windspeeds}. \citet{2020MNRAS.499..873S} studied the mass loss rates and wind speeds of helium stars, with a suite of hydrodynamically consistent atmosphere models. Following the results of \citet{2020MNRAS.499..873S}, where log$_{10}$(L/L$_{\odot}$)$\propto$0.3$\times$log$_{10}$(V$_\mathrm{wind}$/V$_\mathrm{esc}$), we use the \citet{2000A&A...360..227N} WC prescription for all Wolf-Rayet stars, as this best approximates these next generation atmospheric modelling results. The exception to this is at the (metallicity dependant) lower luminosity limit of Wolf-Rayet behaviour \citep{2020A&A...634A..79S}. 

\begin{figure*}
\centering
\includegraphics[width=0.99\textwidth]{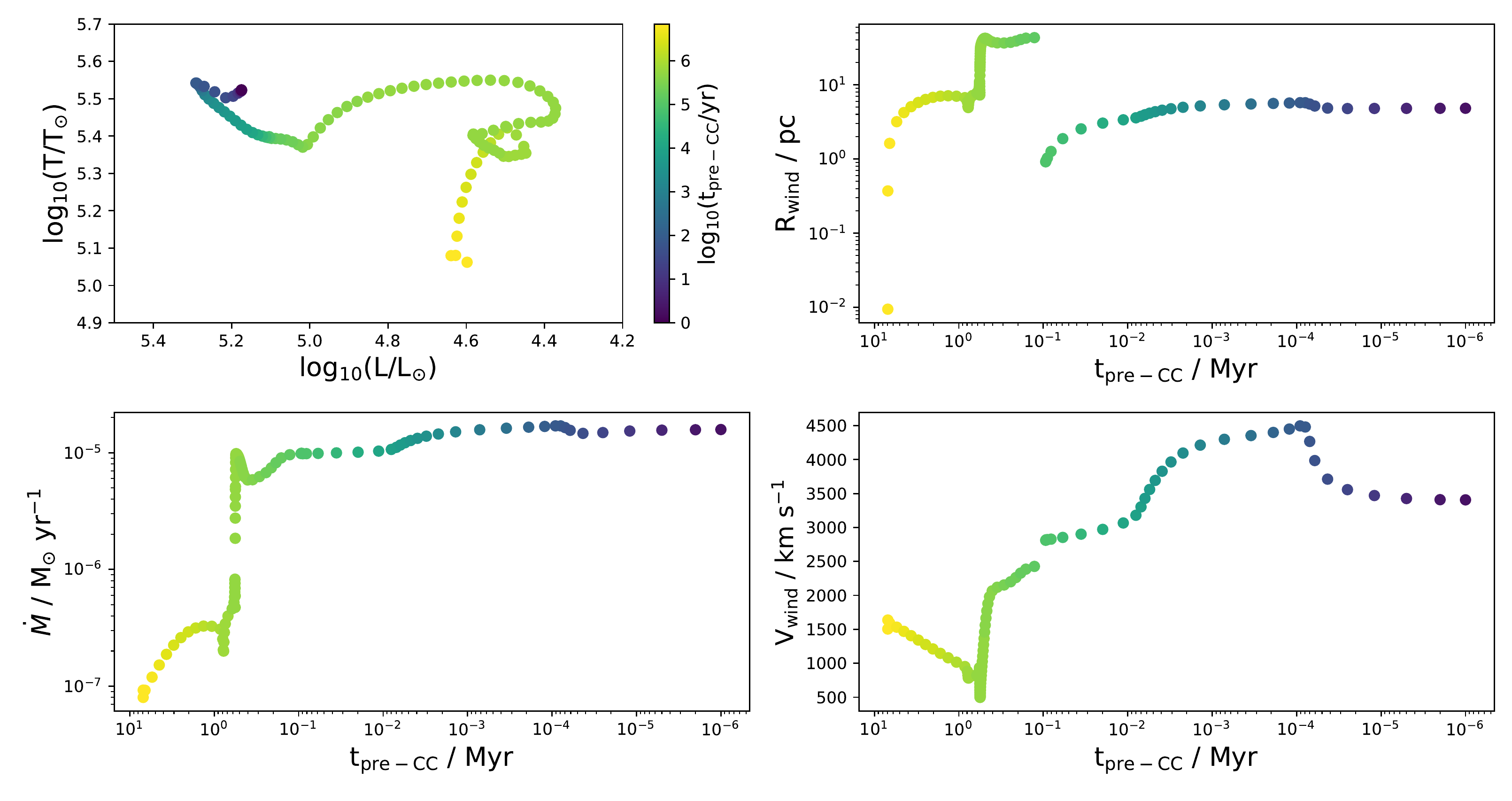}
\caption{An example of the analytic solutions of \citet{1995ApJ...455..145G} applied to the BPASS models. A z002-30-0.6-0.6 model is shown (0.1Z$_{\odot}$, 30M$_{\odot}$ primary, mass ratio of 0.6 and log$_{10}$(P/days)=0.6) with an ISM density $n=$1\,cm$^{-3}$. $r_\mathrm{wind}$ here is the innermost bubble radius - when the Wolf-Rayet phase starts at ${\sim}$0.1\,Myr before core-collapse, the radius works back up from 0. We show evolution in the HR diagram (stellar temperature versus luminosity, top left), wind termination shock radius (top right), mass loss rate (bottom left) and wind speed (bottom right) as a function of time until core-collapse (CC, not exact since the BPASS models stop at the end of core carbon burning). The mechanical wind luminosity at late times, given by Equation \ref{eq:mwl}, is $\sim 5 \times 10^{37}$\,erg\,s$^{-1}$.}
\label{fig:example}
\end{figure*}

\subsection{Analytic bubble prescription and application to BPASS models}
\citet{1995ApJ...455..145G} provide analytic solutions for a Wolf-Rayet wind blowing into a main-sequence or supergiant wind. There are two extremes that can be considered: the bubble expansion is driven solely by thermal pressure \citep[as assumed by][]{1977ApJ...218..377W}, or it is momentum driven (where the shock has cooled). If external pressure from a hot ISM is considered, the problem becomes analytically intractable. \citet{1995ApJ...455..145G} build on the \citet{1977ApJ...218..377W} model, by considering three distinct wind phases. These are the main sequence, supergiant/LBV, and Wolf-Rayet (corresponding to low density/high velocity, high density/low velocity, and a high density/high velocity wind respectively). \citet{1995ApJ...455..160G} and \citet{1996A&A...316..133G} perform dynamical simulations to check the validity of their analytic solutions, and find that although they do not capture physics such as instabilities in the outflow, the 3-wind model is a good approximation for most massive star evolutionary paths.

At each time-step of the BPASS models, mass loss rate, luminosity, total stellar mass, stellar radius, effective temperature and chemical composition are given as outputs. Wind speeds can be calculated from these parameters as described in the previous section. Working with the thin shell approximation of \citet{1977ApJ...218..377W}, this states that the wind speed is supersonic, the ISM is cold with negligible thermal pressure, and that the shocked wind is adiabatic. The bubble's expansion is therefore driven by the thermal pressure of the shocked stellar wind \citep{1995ApJ...455..145G}. The mechanical wind luminosity is given by,
\begin{equation}\label{eq:mwl}
    L_\mathrm{w} = \frac{1}{2}\dot{M_\mathrm{w}}V_\mathrm{w}^{2},
\end{equation}
where a running mean of the mechanical wind luminosity is tracked as the stellar model evolves;
\begin{equation}
    \bar{L_\mathrm{w}} = \frac{\sum L_\mathrm{w}(t<t_\mathrm{elapsed})}{t_\mathrm{elapsed}}
\end{equation}
and the wind termination shock radius at each time step is given by
\begin{equation} \label{eq:rsgwind}
    r_\mathrm{wind} = \Big(\frac{\dot{M_\mathrm{w}}V_\mathrm{w}}{28}\Big)^{\frac{1}{2}} \Big(\frac{3850}{\bar{L_\mathrm{w}}}\Big)^{\frac{1}{5}} t^{\frac{2}{5}}(\pi \rho_{0})^{-\frac{3}{10}},
\end{equation}

where $\rho_{0}$ is the ISM density. If the star transitions into a Wolf-Rayet, $r_\mathrm{wind}$ resets, and we follow the expansion of the new Wolf-Rayet blown bubble into the CSM produced by the previous phases. An `A$_{0}$' quantity is tracked, representing the density of the wind. This quantity is tracked for any post-main sequence, pre-WR phase where the wind is slow and dense. It is defined as,
\begin{equation}
    \mathrm{A_{0}} = \rho_{0}r_{0}^{2} = \frac{\dot{M_\mathrm{0}}}{4\pi V_\mathrm{0}}
\end{equation}
and the inner termination shock in this dense wind lies at,
\begin{equation}\label{eq:rw}
    r_\mathrm{wind} = \Big(\frac{\dot{M_\mathrm{w}}V_\mathrm{w}}{4\pi\times\mathrm{A_{0}}}\Big)^{\frac{1}{2}}t
\end{equation}
where a time-weighted average of the A$_{0}$ parameter since leaving the main sequence is used. The wind density parameter \astar\ is calculated at each time step and is defined as,
\begin{equation}\label{eq:astar}
    A = \frac{\dot{M}}{4\pi V_{w}} = 5 \times 10^{11} A_{\star} \mathrm{g\,cm^{-3}}
\end{equation}
where $A = \rho r^{2}$ \citep{2000ApJ...536..195C}. \astar\ is a normalisation of the wind density, for $\dot{M}=10^{-5}M_{\odot}$yr$^{-1}$ and $V_{w}=1000$kms$^{-1}$. If there is no Wolf-Rayet phase, $A = A_{0}$.

\begin{figure}
\centering
\includegraphics[width=0.99\columnwidth]{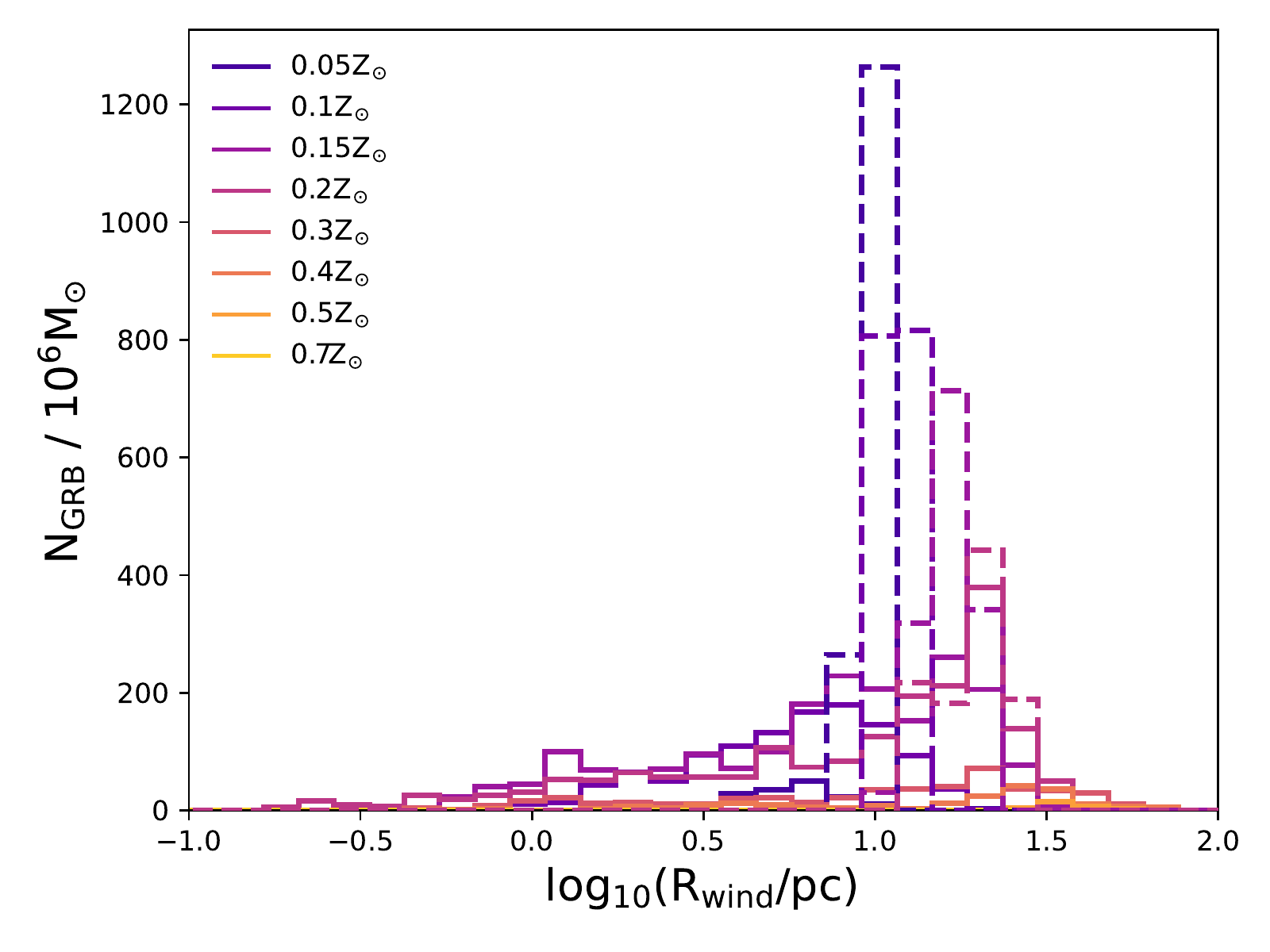}
\caption{The distribution of final termination shock radii $r_\mathrm{wind}$ across all metallicities and models, using the analytic method, at $n=1$\,cm$^{-3}$. The results are coloured by metallicity. QHE models are shown with dashed lines, tidal pathways with solid lines.} 
\label{fig:Rwind}
\end{figure}

\begin{figure}
\centering
\includegraphics[width=0.99\columnwidth]{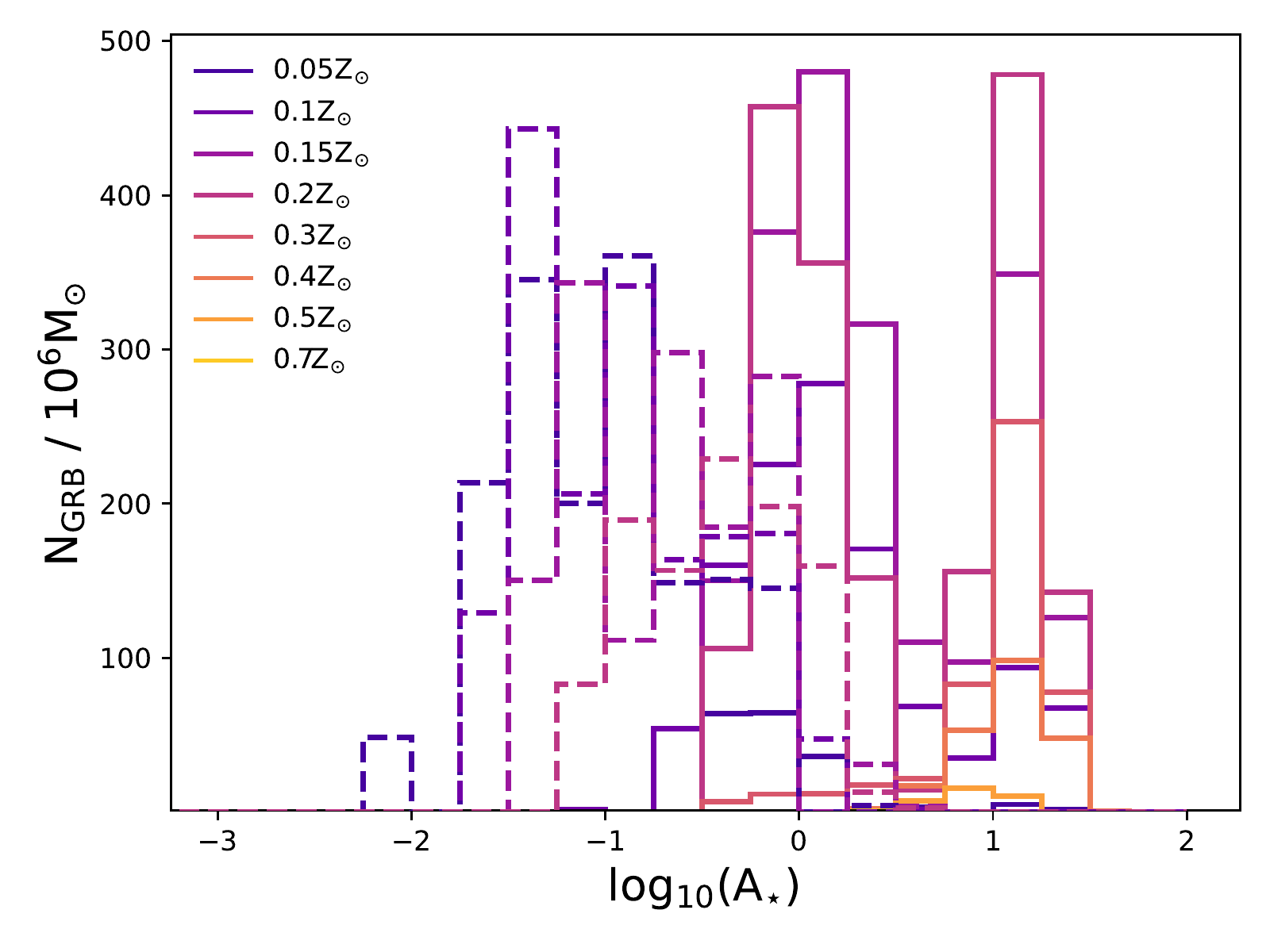}
\caption{The \astar\ distribution for all metallicities considered. Tri-modality can be seen, with QHE models (dashed lines) dominating at low \astar. The low metallicity models have weaker winds, which would produce smaller bubbles, but this is offset by their longer lifetimes (see Figure \ref{fig:DTD}), giving the bubbles longer to expand. The result is that low metallicity models do not dominate the lowest wind radii (Figure \ref{fig:Rwind}). The other two peaks correspond to non-Wolf-Rayet and Wolf-Rayet tidal models.}
\label{fig:Astar}
\end{figure}

\begin{figure}
\centering
\includegraphics[width=0.99\columnwidth]{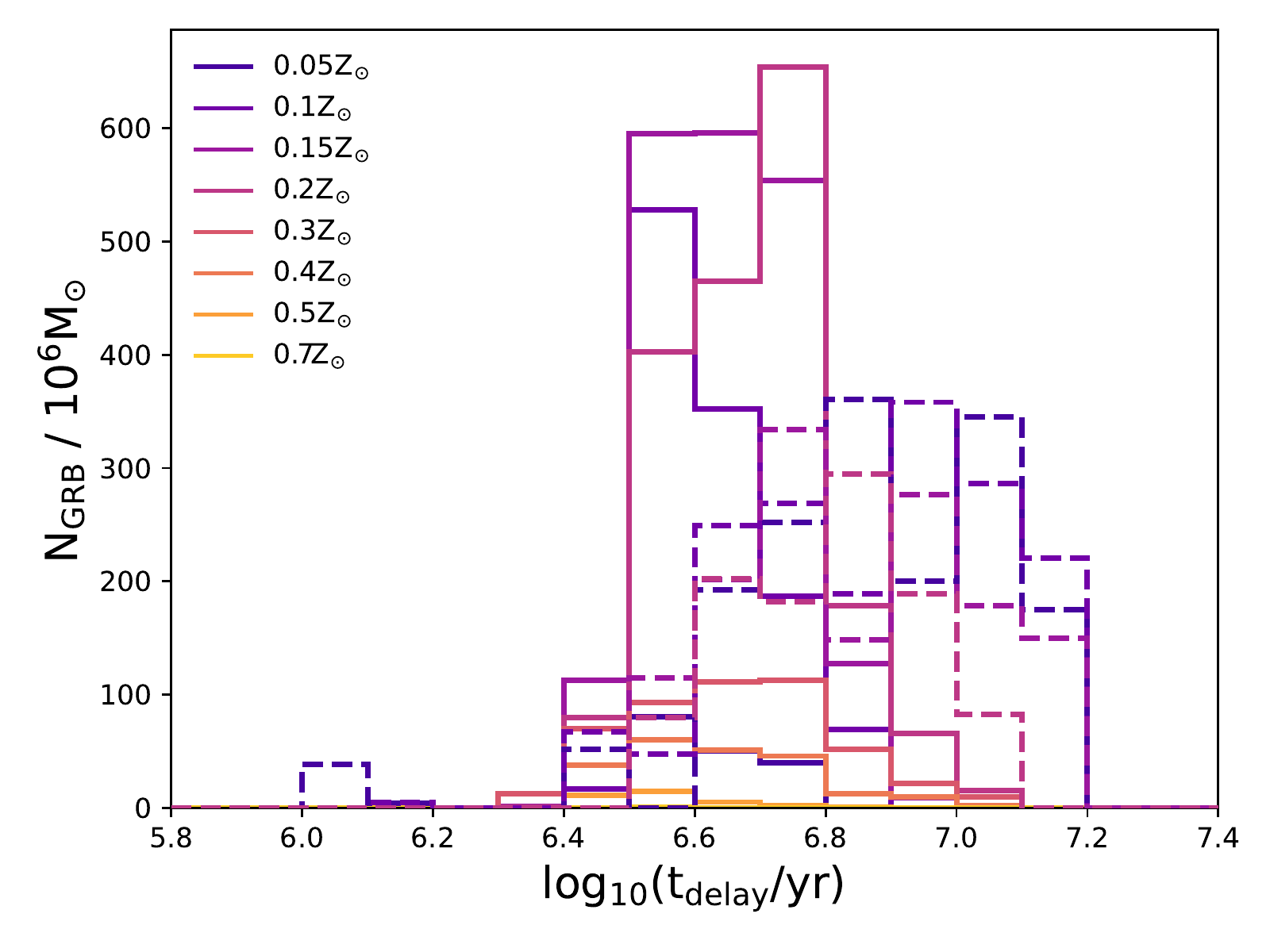}
\caption{Delay time distributions for the progenitor models. The low-Z QHE models (dashed lines) typically have the longest delay times, while for the non-QHE progenitors (solid lines), higher metallicity models live shorter lives. This is a model selection effect - lower mass and longer lived high $Z$ models typically lose too much angular momentum through winds to be viable long GRB progenitors, assuming strong core-envelope coupling \citep{2020MNRAS.491.3479C}.} 
\label{fig:DTD}
\end{figure}

We note that the final moments of stellar evolution (post carbon burning, where the BPASS models end) are often marked by rapidly varying mass loss \citep[e.g.][]{2007Natur.447..829P}. This can result in the formation of circumstellar shells. However, because this typically occurs within the final few years of the stellar evolution, such shells have no time to reach the wind termination shock and will be swept away in the initial explosion. We explore this further in the discussion.

\subsection{Wind bubble population synthesis}
In order to produce predictions for $r_\mathrm{wind}$, we run these analytic calculations for every LGRB progenitor model at a range of ISM densities, from $n=10^{-3}$ to $10^{7}$\,cm$^{-3}$, with order of magnitude steps, for a total of 11 trial densities.

The resultant distributions of \astar\ and $r_\mathrm{wind}$ are weighted by the BPASS weightings, which are informed by the observed distribution of binary parameters in the Galaxy \citep{2017ApJS..230...15M}, and a Kroupa IMF up to a maximum mass of 300\,M$_{\odot}$ \citep{2001MNRAS.322..231K}. Since the models have been selected by matching the observed LGRB rate, and the metallicity distribution that arises is comparable to the observed host distribution (the same population for which we have modelled the afterglows in Section \ref{sec:obsv}), no further metallicity weighting is required to match the observed sample.

Our analytic $r_\mathrm{wind}$ results are shown in Figure \ref{fig:Rwind} for $n = 1$\,cm$^{-3}$. There is an approximately equal contribution from QHE and binary tidal pathways. The distribution has a strong peak, in part due to the QHE models, for which there are a limited number of BPASS models but which contribute significantly to the rate \citep[resulting in a small number of models, and hence radii and wind densities, being heavily weighted,][]{2017PASA...34...58E}. Furthermore, Equation \ref{eq:rw} demonstrates that $r_{\rm wind}$ has a weak dependence on the wind parameters, also decreasing the spread in $r_{\rm wind}$ values.

In Figures \ref{fig:Astar} and \ref{fig:DTD} we show the \astar\ and delay time distributions for the progenitors. Despite QHE stars having the lowest metallicites (and weakest winds), they do not dominate the lowest wind termination radii (Figure \ref{fig:Rwind}). This is because they tend to have longer lifetimes (by a few Myr, or a factor of two), as shown in Figure \ref{fig:DTD}, giving more time for bubble expansion. Conversely, the higher metallicity models have shorter lifetimes. This is not intrinsic to the models, but a selection effect: high $Z$ and long-lived stars lose too much mass and angular momentum to be viable long GRB progenitors \citep{2020MNRAS.491.3479C}. As the QHE and low-$Z$ winds are weaker by around a factor of 10 in \astar\, and the wind radius approximately scales as the square-root of \astar\, the lifetime differences approximately cancel to yield QHE and low-$Z$ wind radii that are comparable.

\section{Exploring the effect of assumptions in the analytic model}\label{sec:hydro}
\subsection{Hydrodynamical simulations}
The analytic solutions of \citet{1977ApJ...218..377W} and \citet{1995ApJ...455..145G} naturally incorporate several assumptions. In order to gain an understanding of how the analytical solutions for $r_{\rm wind}$ found in Section \ref{sec:analytic} deviate from the results obtained from more physically motivated hydrodynamical simulations, we make use of the PLUTO code \citep{2007ApJS..170..228M,2012ApJS..198....7M,2014JCoPh.270..784M}. The code solves a system of integrated conservation laws, where the fluid inflow, outflow, temperature and density are tracked in each cell of the simulation. We set up a structured mesh with the 'HD' module, which uses pure Newtonian hydrodynamics, neglecting magnetic fields and special relativistic effects. The latter is a safe assumption at $\sim$1000kms$^{-1}$ wind speeds and temperatures no higher than 10$^{5}$ K, while the lack of magnetic fields is primarily to avoid introducing another variable. Their potential impact is discussed in Section \ref{sec:discuss}.

Our simulations are set up as follows. We use a Runge-Kutta-Legendre (RKL) time-stepping scheme, accurate to second order in time, without dimensional splitting due to our adoption of a cylindrical geometry. Short timescale processes (such as radiative cooling) can then be well approximated, albeit with increased computational cost \citep[see e.g.][]{2009ApJS..181..391T}. Cooling is included with an explicit scheme, where the rate of energy loss due to cooling in each cell is proportional to $n^{2}\Lambda(T)$, and we have used the cooling curves of \citet{2009A&A...508..751S}. Full ionisation is assumed above 10,000\,K and an ionisation fraction of $10^{-3}$ below. The temperatures in our simulations are typically of this order of magnitude. With cooling included, the stellar wind shock can be radiative, and thermal pressure becomes less important relative to the momentum of the wind. The CFL number is limited to CFL$\leq 0.4$, and a hll Riemann solver is used \citep{2014JCoPh.270..784M}.

\citet{2009A&A...508..751S} provide the cooling constant ${\Lambda}$ for Solar metallicity as a function of temperature, and also the contributions from each element at Solar metallicity so that the cooling curve can be adjusted for non-Solar abundances. The curves are scaled by adjusting the contribution from the available elements, according to the relative BPASS abundances at each metallicity \citep{2017PASA...34...58E}. Because not every element calculated by \citet{2009A&A...508..751S} is available in the BPASS outputs, we only scale the elements that are in BPASS, and keep the others fixed at the Solar value from the fiducial \citet{2009A&A...508..751S} curve. These are in every case trace elements (e.g. Na, Ar) that do not significantly contribute to cooling. We run the simulations at ZAMS (zero age main sequence) metallicity and do not adjust this throughout the evolution. This is primarily an issue for Wolf-Rayet phases, where the hydrogen abundance of the outflow is reduced, and heavier elements are enhanced. We will discuss the effect of this approximation in Section \ref{sec:discuss}.

To cater for the rare cases where the simulation temperature exceeds the range provided by \citet{2009A&A...508..751S}, we linearly extrapolate the upper end of the curve at each metallicity out to 5${\times}$10$^{9}$ K. This approximates the high temperature, Bremsstrahlung-dominated cooling trends calculated by \citet{2020MNRAS.tmp.2268P}. We do not employ those curves directly due to significant uncertainties over the exact parameters (e.g. radiation field, dust content) to use, and varying these additional parameters is beyond the scope of this work. We also exclude the effects of dust in our cooling. However, for high-temperature stars at low metallicity, dust formation is likely to be somewhat limited \citep{2010ApJ...713..592E}. No thermal conduction and a simple ionisation structure are assumed. For detailed study of the effect of these, we refer the reader to e.g. \citet{2011ApJ...737..100T}.

Our wind simulation strategy is as follows. A stellar wind with a $r^{-2}$ profile is initialised within a semicircular region of radius 2\,pc (8 or 16 cells, depending on the resolution chosen). Given our assumption of spherical symmetry, this yields equivalent results at half the computational cost. The full simulation region is fixed at either $300\times600$ or $600\times1200$ cells, with 0.25\,pc\,cell$^{-1}$ or 0.125\,pc\,cell$^{-1}$. 
We run the code in a highly parallelised configuration. Unit quantities are chosen to be 1\,pc, 1 proton mass per cm$^{3}$, and 10$^{8}$ cms$^{-1}$, with cm, seconds and grams as the units. The unit time step is therefore 970 years. The {\sc init.c} file used to initialise the simulations is based on the stellar wind example available with PLUTO \citep{2014JCoPh.270..784M}. The wind is parameterised as ${\rho}\propto Sr^{-2}$, where $S$ is scaled so that the wind has the correct density profile for the given $\dot{M}$, $V_\mathrm{wind}$ and simulation units. 

The stellar evolution is split into 2-4 phases: main sequence, RSG/YSG/BSG, eruptive episodes (if this occurs and is distinct from other phases) and Wolf-Rayet (if this occurs). The mean mass loss rate and wind velocity are averaged over the duration that the star spends as a specific spectral type (OBAFGKM or Wolf-Rayet), and are used to initialise the simulation wind profile at the start of each stage. The final density, pressure and velocity fields from the previous stage are used as the initial conditions for the next, upon which the new wind profile is superimposed, within 2\,pc of the origin, before the simulation is restarted. We are careful not to create scenarios where the superimposed wind in the first stage (when there is no existing simulated bubble, just an ISM) is larger than the expected termination shock radius - this is done by estimating $r_\mathrm{wind}$ from the analytic solution.

\subsection{Verification of simulations}\label{sec:verify}
While the detailed simulations described above contain physics which are not accounted for in the analytic model, if we simulate the conditions assumed in the analytic model, we should obtain a similar result. To test this and verify the reliability of the simulations, we run a 60\,M$_{\odot}$ single star model at $0.4Z_{\odot}$ for 3.5\,Myr, in an $n=100$\,cm$^{-3}$ (100K) and $n=1$\,cm$^{-3}$ (8000K) ISM. This model is not selected as a GRB progenitor, but is used here as a standard comparison model with others studied in the literature \citep[e.g.][]{2006MNRAS.367..186E,2011ApJ...737..100T,2012A&A...547A...3V}. 

\begin{figure*}
\centering
\includegraphics[width=0.50\textwidth]{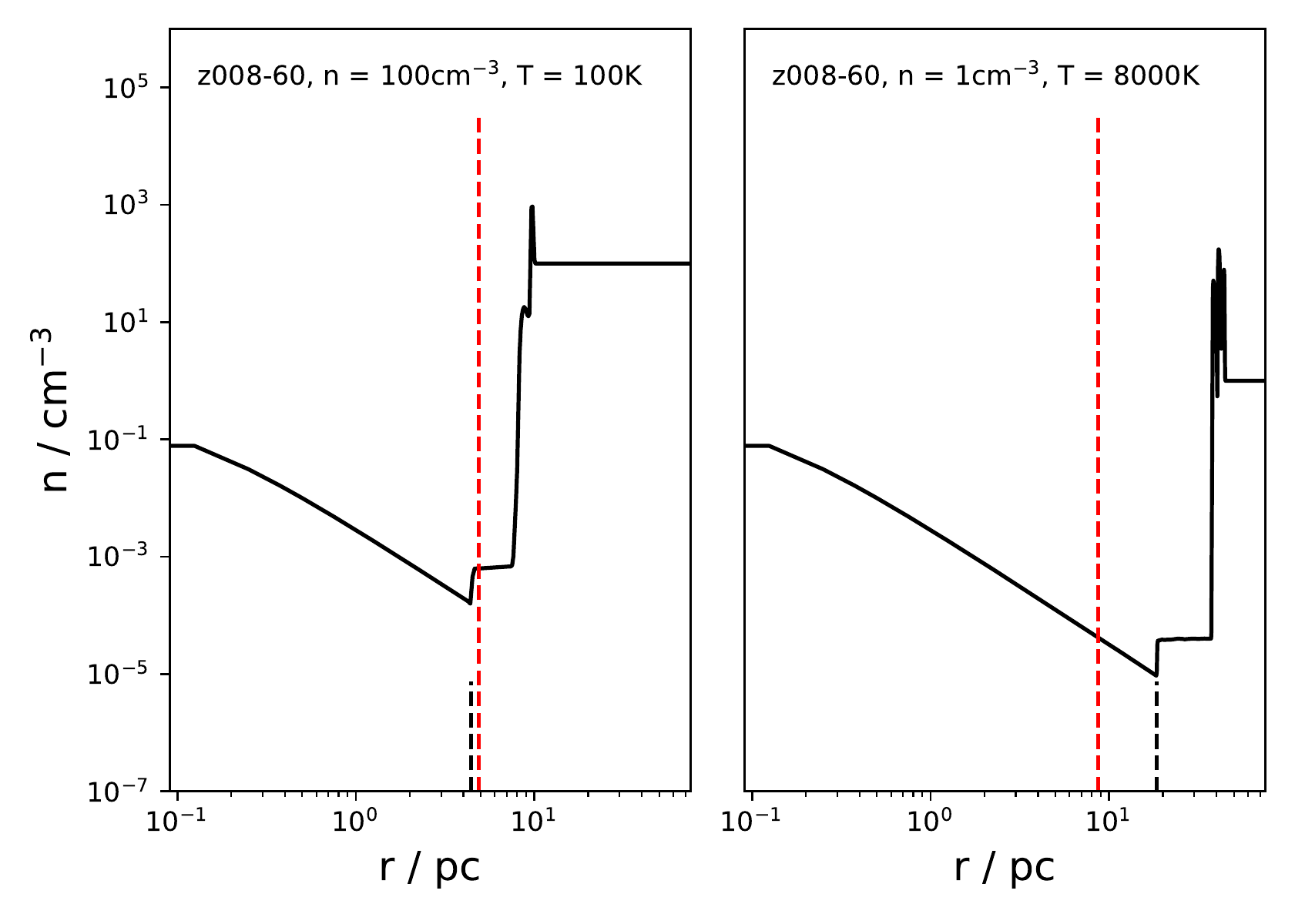}
\includegraphics[width=0.48\textwidth]{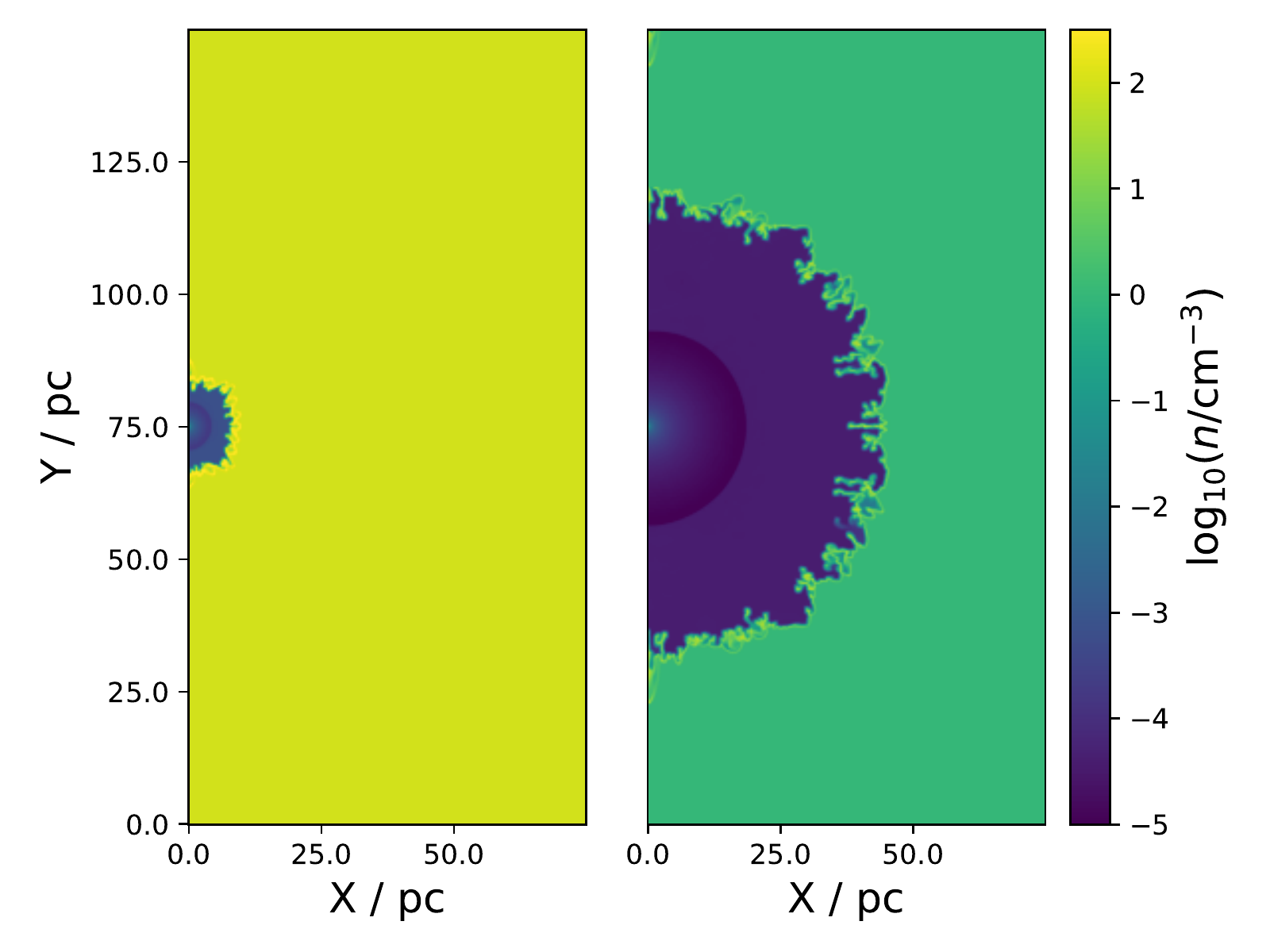}
\caption{Test simulations performed with {\sc PLUTO}. In each case, the $r^{-2}$ wind is initialised in a region of radius 2\,pc centered at (0,75). The wind parameters are based on a $0.4Z_{\odot}$, $60M_{\odot}$ BPASS model, chosen as the test example for its similarity to simulations performed in the literature \citep[e.g.][]{2006MNRAS.367..186E,2011ApJ...737..100T,2012A&A...547A...3V}.
The model has a mean main sequence mass loss rate of log$_{10}$($\dot{M}/M_{\odot}$yr$^{-1}$)$=-5.73$ and a mean wind speed of 1740\,kms$^{-1}$. The simulation is stopped at 3.5\,Myr (shortly before the end of the main sequence). The first run uses an ISM density of 100\,cm$^{-3}$ and a cold ISM of $T=100$K, the second uses 1\,cm$^{-3}$ and $T=8000$K. The two resultant 1D profiles are shown in the left two panels, and the corresponding 2D density maps on the right. The 1D profiles are drawn from $y=75$\,pc. As expected, the dense/cold example reproduces the \citet{1977ApJ...218..377W} $r_{\rm{wind}}$ prediction well (4.5 versus 4.9\,pc), whereas the low density/hot ISM produces 18.6 versus 8.7\,pc, a factor of $\sim$2 difference. In the 1D plots, the analytic r$_{\rm wind}$ values are indicated by dashed vertical red lines, the position of the hydrodynamical termination shocks are noted by shorter dashed black lines. }
\label{fig:verification}
\end{figure*}

The results are shown in Figure \ref{fig:verification}. In the cold/dense example, the termination shock radius in the simulation is 4.5\,pc, versus 4.9\,pc from the analytic model. This is expected because the models of \citet{1977ApJ...218..377W} and \citet{1995ApJ...455..145G} assume (i) an infinitely thin outer shock, and (ii) negligible thermal pressure from the ISM. They also assume no energy loss due to cooling. 
A dense, cold ISM satisfies criteria (i) and (ii). Aided by our inclusion of radiative cooling in the simulations, which is more efficient at high densities, the outer shock is compressed further.

However, when a hot, low density ISM is used, the assumptions of the analytic model break down - we find 8.7 versus 18.6, a factor of 2.1 difference. The shock is less compressed and has a lower temperature. Since the thermal pressure in the shock and stalled wind must match the wind ram pressure for the bubble to be in equilibrium, and ram pressure decreases with distance from the star, a less compressed outer shock results in a greater radius at which equilibrium is reached, with respect to a model assuming a thin outer shell. Furthermore, if the bubble expands faster than the sound-crossing timescale of the shocked wind, the semi-static assumption of the analytic models also breaks down. Overall, this demonstrates that the analytic method can be inaccurate in ISM temperature and density regimes that are entirely plausible for GRB progenitors.

\subsection{Corrections to the analytic results}
We used analytic solutions to make a population study feasible in Section \ref{sec:analytic}, as running detailed hydrodynamical simulations for hundreds of models would be unreasonably computationally expensive. However, the more physically motivated simulations yield different results, as shown in the previous section. Since running hydrodynamical simulations for every model is a significant (and perhaps unnecessary) undertaking, we instead consider the application of corrections to the analytic results for $r_{\rm wind}$ to better match hydrosim results. This is done by performing simulations for 16 combinations of progenitor model and ISM density, as listed in Table \ref{tab:hydrosims}, calculating the factor increase in $r_{\rm wind}$ in the simulation over the analytic result, and finding trends in this increase such that a representative correction can be applied to any given analytic result.

\begin{table}
\centering 
\caption{Models for which have performed hydrodynamical simulations. The first two are z002-50-0.6-0.6 (0.1Z$_{\odot}$, 50\,M$_{\odot}$ primary, 0.6 mass ratio and log$_{10}$(P/days)$=0.6$) and z004-50hmg (a 50\,M$_{\odot}$ QHE model at 0.2Z$_{\odot}$). These are both evaluated at $n=$ 0.05, 0.1, 1, 10 and 100\,cm$^{-3}$. A z014-60-0.7-0.7 model is simulated at 0.1, 1 and 10\,cm$^{-3}$, and a z010-60-0.9-0.0 model at 0.05 and 10\,cm$^{-3}$. Additionally, a single star 60\,M$_{\odot}$ model (not a GRB progenitor) is run at 1\,cm$^{-3}$ for verification purposes (Section \ref{sec:verify}). For each, we list the analytic prediction for the termination shock radius $r_{\rm wind}$ and the hydrosimulation equivalent, the radius of the outer edge of the stalled wind $R_{\rm sw}$, and the average density log$_{10}(n_{\rm sw})$ across the stalled wind region.} 
\label{tab:hydrosims}
\begin{tabular}{llllll}
\hline %
Model & ISM $n$ & $R_{\rm{wind,a}}$ & $R_{\rm{wind,h}}$ &  $R_{\rm{sw}}$ & log$_{10}$($n_{\rm{sw}}$) \\
\hline %
z002-50-0.6-0.6	&	1	&	8.69	&	21	&	39.7	&	-3.93	 \\
z002-50-0.6-0.6	&	0.1	&	13.77	&	33	&	>75	&	-4.34	 \\
z002-50-0.6-0.6	&	10	&	5.48	&	9.7	&	20	&	-3.24	 \\
z002-50-0.6-0.6	&	100	&	3.97	&	4.06	&	9	&	-2.49	 \\
z002-50-0.6-0.6	&	0.05	&	15.81	&	54.9	&	>75	&	-4.74	 \\
z004-50hmg	&	10	&	9.69	&	15.56	&	28.5	&	-3.74	 \\
z004-50hmg	&	1	&	15.36	&	31.27	&	52	&	-4.27	 \\
z004-50hmg	&	0.1	&	24.34	&	42.13	&	>75	&	-4.53	 \\
z004-50hmg	&	0.05	&	27.96	&	52.54	&	>75	&	-4.73	 \\
z004-50hmg	&	100	&	6.11	&	8.22	&	14.45	&	-3.13	 \\
z014-60-0.7-0.4	&	1	&	7.56	&	38.9	&	62	&	-3.87	 \\
z014-60-0.7-0.4	&	0.1	&	11.99	&	63.38	&	>75	&	-4.26	 \\
z014-60-0.7-0.4	&	10	&	3.05	&	21.8	&	33.9	&	-3.38	 \\
z010-40-0.9-0	&	0.05	&	9.86	&	63.8	&	>75	&	-4.49	 \\
z010-40-0.9-0	&	10	&	3.42	&	21.5	&	33.1	&	-3.69	 \\
z008-60	&	1	&	12.27	&	27.9	&	43	&	-3.92	 \\
\hline %
\end{tabular}
\end{table}


\begin{figure*}
    \centering
    \includegraphics[width=\textwidth]{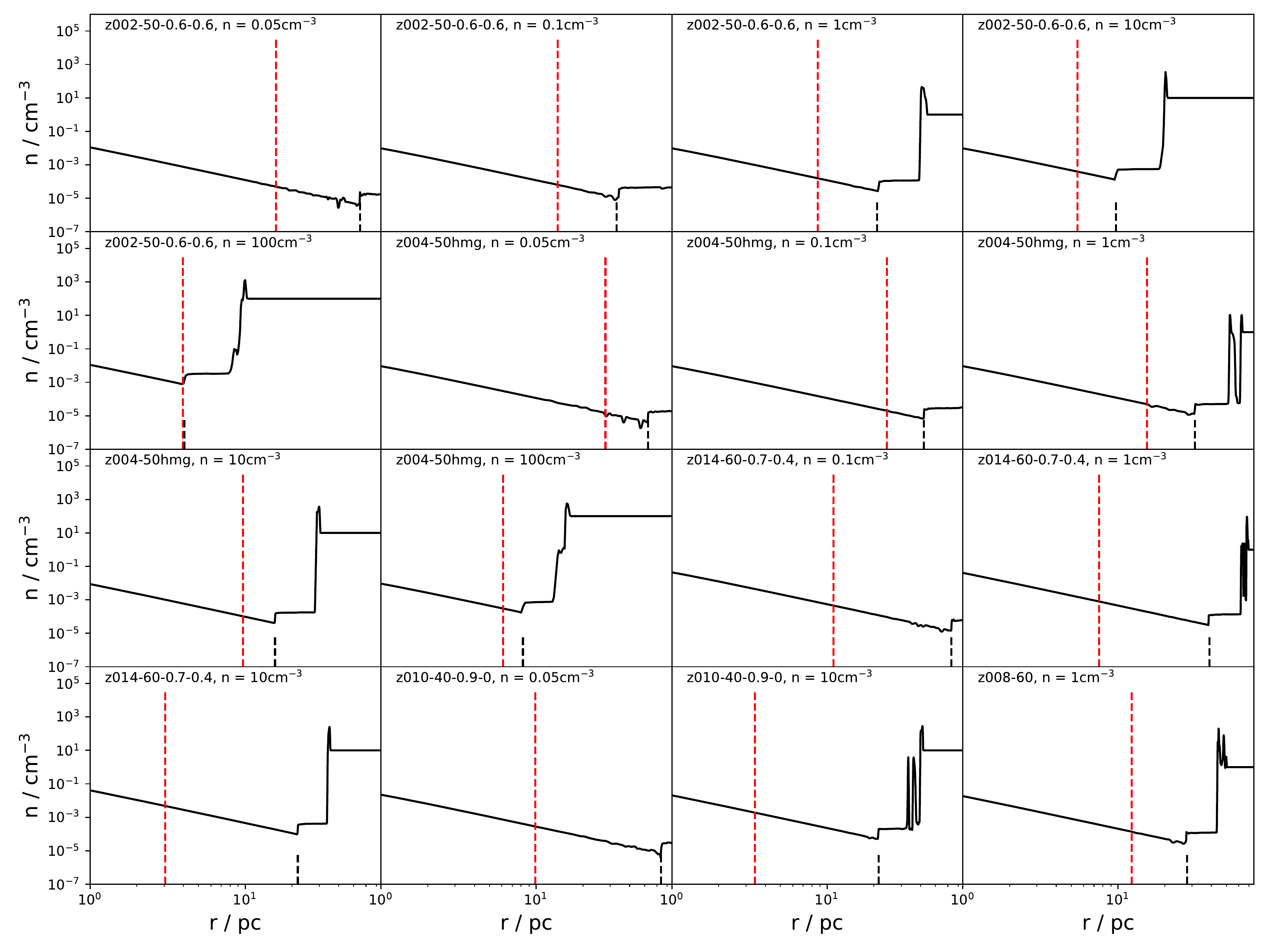}
    \caption{The 16 1D CSM profiles pre-collapse, from simulations that were run in order to (i) compare the result for $r_\mathrm{wind}$ to their analytic counterparts and (ii) evaluate the range of stalled wind densities that are possible for a fixed ISM density. The model used and ISM density are indicated in each case, and the results listed in Table \ref{tab:hydrosims}. The final analytic values for $r_\mathrm{wind}$ are indicated by a dashed red lines, the hydrodynamical values by shorter dashed black lines.  }
    \label{fig:hydrosims}
\end{figure*}

\begin{figure*}
\centering
\includegraphics[width=0.99\textwidth]{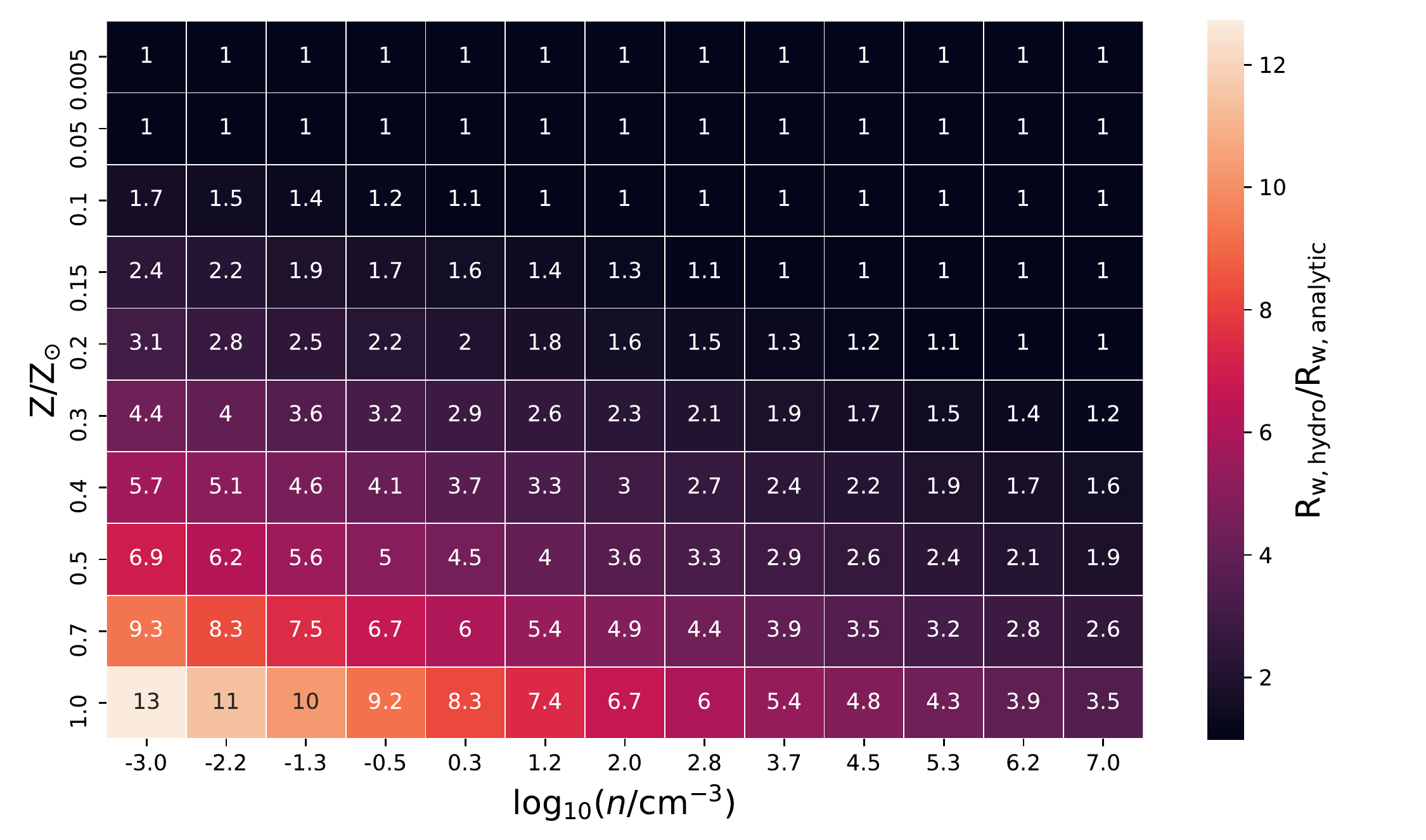}
\caption{The factor by which $r_\mathrm{wind}$ increases in a hydrosimulation, versus the analytic result using the same stellar evolution model. Only a small proportion of the hydrosimulation values in the grid were performed, the gaps were filled by fitting functions and interpolating (see text and hydrosimulation outputs in Figure \ref{fig:hydrosims}). The corrective factor tends to 1 at low $Z$ and high density, likely because the cooling difference between the approaches is reduced, and a dense ISM better reproduces the thin shell structure assumed in the analytic model. We have floored the corrections at 1.}
\label{fig:upscatter}
\end{figure*}

The resultant 1D density profiles are shown in Figure \ref{fig:hydrosims}. In each case, the analytic model prediction for $r_\mathrm{wind}$ is indicated by a vertical dashed line. The hydro results for $r_\mathrm{wind}$ are consistently higher than the analytic results. We assume a priori that the metallicity and ISM density will be important factors in determining the deviation from the analytic $r_{\rm wind}$, since these affect cooling and the compression of the outer shock. Fitting power laws across this $Z$-$n$ parameter space, we find that the factor increase in $r_\mathrm{wind}$ over the analytic result can be described by $270n^{-0.056}Z^{0.88}$, where $Z$ is the metallicity by mass fraction and $n$ is the ISM density in cm$^{-3}$. For those models which have both analytic and hydrodynamic results, the standard deviation between the corrective factor from this fit and the hydro/wind ratio in each case is 0.4. There is therefore considerable scatter in the relation, but it nevertheless approximates the results that would have been obtained, had we performed simulations for each. The corrections across $Z$-$n$ parameter space are shown in Figure \ref{fig:upscatter}, extrapolated down to $10^{-3}$ and up to $10^{7}$\,cm$^{-3}$. The $r_\mathrm{wind}$ corrective factor is floored at 1 (i.e. we never decrease it below the analytic value).

This approach of comparing analytic to hydrodynamical results has been performed before, but not for a whole suite of binary models. \citet{2006MNRAS.367..186E} ran a selection of numerical simulations to determine the range of $r_\mathrm{wind}$ and \astar\ for Wolf-Rayet stars. They found a typical difference of ${\sim}$0.3 dex between the analytic and numerical results. At $n=1$\,cm$^{-3}$, we also find modest increases of ${\sim}$0.3 dex at low metallicity. At higher metallicity we find a more significant deviation (a few times greater); this may be due to our inclusion of radiative cooling - more impactful at high metallicity - but not applied in the simulations of \citet{2006MNRAS.367..186E}.

\begin{figure}
\centering
\includegraphics[width=0.99\columnwidth]{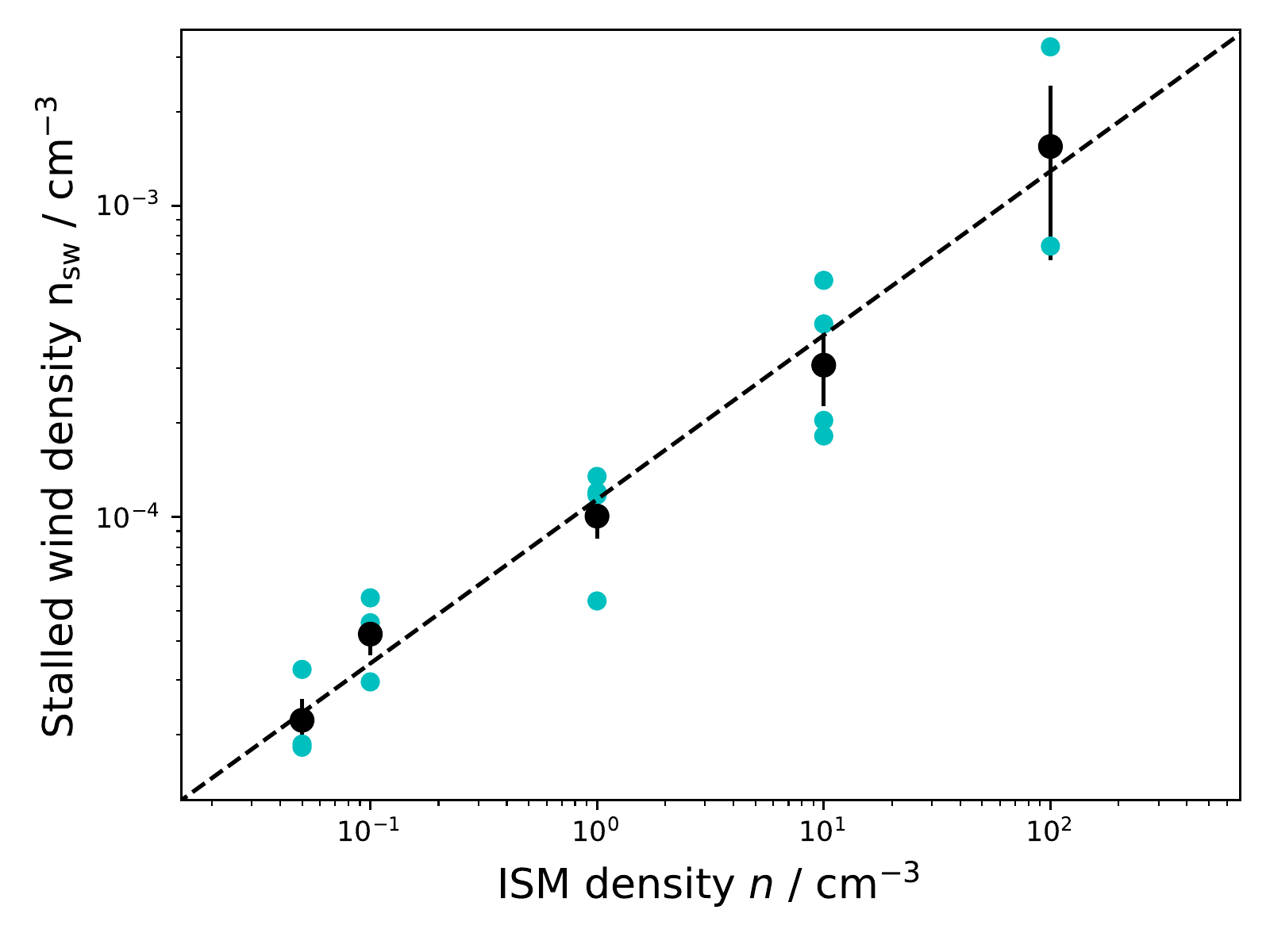}
\caption{The stalled wind density, for the same models as shown in Figure \ref{fig:hydrosims}, versus the ISM density. Cyan dots are individual results, larger black dots are means, with the standard error on the mean indicated. Over the range of $n$ simulated, there is a linear relation in log-log space. The stalled wind has a flat density profile and may produce ISM-like afterglow evolution at lower radii than the ISM itself.}
\label{fig:swdensity}
\end{figure}

A central question is whether the complex shock structure, including a stalled (shocked) wind region beyond $r_{\rm wind}$ but before the ISM, can help explain observations of low constant density media. In addition to correcting the analytic termination shock radius, we also determine the mean outer radius of the stalled wind (i.e. the shock boundary between the stalled wind and ISM) in the hydrosimulations, $r_{\rm ism}$, to be 1.77$\pm$0.07 times the hydrosim termination shock. Within the stalled wind, the densities are approximately constant (see Figure \ref{fig:hydrosims}), and follow a linear relation in log-log space with the ISM density, as shown in Figure \ref{fig:swdensity}. We can therefore select, at 11 hours rest-frame, (i) a wind density $A_{\star}$, for $r_{\rm emit} < r_{\rm wind}$, (ii) a stalled wind density for $r_{\rm wind} < r_{\rm emit} < r_{\rm ism}$ or (iii) the ISM density for $r_{\rm emit} > r_{\rm ism}$. This three-region model approximates most of the CSM simulations well. Hence, starting with analytic $r_{\rm wind}$ and $A_{\star}$ values for each BPASS model and ISM density combination, we can obtain density estimates for the three regions by applying analytic-hydrosim corrections. 

To summarise Sections \ref{sec:analytic} and \ref{sec:hydro},
\begin{itemize}
    \item R$_{\rm wind}$ is obtained for each BPASS model from the analytic estimate (Equations \ref{eq:rsgwind} and \ref{eq:rw}) by multiplying a corrective factor, dependent on metallicity $Z$ and ISM density $n$, as shown in Figure \ref{fig:upscatter},
    \item the outer stalled wind radius (interface with the ISM) is obtained from a tight $r_{\rm ism}$ = ($1.77\pm0.07)\times r_{\rm wind}$ relationship,
    \item the stalled wind density in the region $r_{\rm wind} < r < r_{\rm ism}$ is estimated from an $n$-$n_{\rm ism}$ correlation, demonstrated in Figure \ref{fig:swdensity},
\end{itemize}
and $A_{\star}$ is a product of the stellar evolution model parameters (Equation \ref{eq:astar}). All that remains is to decide upon the emission radii with which to sample each model's CSM, and to generate predictions for a variety of ISM densities.

\section{Results}\label{sec:results}

\subsection{Observed afterglows}
Our full sample of fit results are presented in Table~\ref{app:fits}. As noted in Section~\ref{sec:obsv}, while we discourage over-interpretation of individual fits, the ensemble can be used to sample the distributions of the physical parameters found in long GRB environments. We find that for ISM-like events, the mean log\,$(E_{\rm k, iso}/$erg$) = 52.99 \pm 0.91$, while for wind-like events it's log\,$(E_{\rm k, iso}/$erg$) = 53.42 \pm 0.87$. The finding of a higher average energy for wind-like events is consistent with the results of \cite{2018ApJ...866..162G}. Indeed, a Kolmogorov-Smirnov (KS) test indicates a probability of $p = 0.03$ that the $E_{\rm k, iso}$ values from the two environments were drawn from the same distribution ($p = 0.04$ for an Anderson-Darling (AD) test). The significance of this dichotomy is slightly in excess of the $2\sigma$ divide found in \citet{2018ApJ...866..162G} and may be due to the factor $\sim 2$ increase in sample size. We also find that $\theta_o$ displays a quasi-Maxwellian distribution that peaks at $\theta_o \approx 0.1$. Both the KS and AD test find a $p \gg 3\sigma$ separation in $\theta_o$ values derived for wind-like and ISM-like environments, but the wind distribution is very broad. Based on a Hellinger distance of $0.03$, we conclude that the wind posteriors for $\theta_o$ are just returning the prior. No trends or divisions are observed in $p$ or $\epsilon_B$, though $p$ shows a preference for lower values ($p \approx 2$) in both wind-like and ISM-like environments.

\subsection{Comparison to the observed Sample}
In the previous sections, distributions for $r_\mathrm{wind}$ and \astar\ were produced using an analytic method, and corrections to $r_{\rm wind}$ applied to better match results from hydrodynamical simulations. However, the distributions in Figures \ref{fig:Rwind}, \ref{fig:Astar} and \ref{fig:DTD} are for {\it all} BPASS models identified as possible progenitors. This is not the sample we would see observationally, and is not what should be compared to the results of Section \ref{sec:obsv}. We can only measure \astar\ in cases where the emission radius $r_\mathrm{emit}$ of the GRB jet is inside $r_\mathrm{wind}$, and can only measure the stalled wind or ISM densities if the jet has reached that far at the time of observation.

To calculate R$_{\rm emit}$ for each model at standardised time of 11 hours post-burst (rest frame), we adopt the emission radius-time relations of \citet{2005ApJ...633.1018P} and \citet{2018ApJ...866..162G}. In a constant density medium (ISM or stalled wind) we have,
\begin{equation}\label{eq:n}
    r_{\rm emit}(t) = 5.85\times10^{17}\bigg(\frac{1+z}{2}\bigg)^{-\frac{1}{4}}E_{53}^{\frac{1}{4}}n^{-\frac{1}{4}}t^{\frac{1}{4}}
\end{equation}
and for a wind-like medium,
\begin{equation}\label{eq:emit}
    r_{\rm emit}(t) = 3.2\times10^{17}\bigg(\frac{1+z}{2}\bigg)^{-\frac{1}{2}}E_{53}^{\frac{1}{2}}A_{\star}^{-\frac{1}{2}}t^{\frac{1}{2}}
\end{equation}
where $z$ is the redshift, $E=E_{53}10^{53}$\,erg, $n$ is the density in cm$^{-3}$, $A_{\star}$ is the wind density parameter and $t_{\rm obs}$ is the rest-frame time post-burst in hours. For each model, we evolve the emission radius from $t=0$ to $t=11$\,hours, passing through wind, stalled wind and ISM regions, depending on how far $r_{\rm emit}$ reaches in 11 (rest-frame) hours. Due to the relativistic speeds and assuming small angles with respect to the jet axis, cosmological rest-frame times of 11 hours post-burst can correspond to years of jet propagation. The relation used to model the propagation changes upon crossing the termination shock as outlined above. We note that the precise of choice of 11 hours as the standardised time is somewhat arbitrary, but that the emission radius is only weakly dependent on time as shown above, evolving as t$^{0.5}$ or t$^{0.25}$.

The wind density parameter is an output from the analytic calculations for a given stellar evolution model, and the ISM density is a pre-chosen value. The density of the stalled wind is calculated with the relation shown in Figure \ref{fig:swdensity}. The only variables left in equations \ref{eq:n} and \ref{eq:astar} are the redshift and burst energy. We assume a priori that these are not intrinsic to the models or clearly related to the ISM density, so we randomly draw values from the redshift and energy distributions of the observational sample (see Section \ref{sec:obsv}, and Section \ref{sec:testrwindeiso} where we investigate the impact of this assumption). We impose a minimum prompt fluence cut for detection of $2\times 10^{-6}$\,erg\,cm$^{-2}$, derived from the randomly drawn E$_{\rm iso}$ and $z$ \citep[luminosity distance,][]{2006PASP..118.1711W}, based on the conservative lower limit among detected GRBs presented by \citep{2019MNRAS.488.5823L}. An efficiency of one is assumed. If the fluence is too low, new E$_{\rm iso}$ and $z$ pairs are continually drawn until an `observable' fluence is selected. We then move on to the next model and repeat the process. In this way, we populate the observable redshift-flux parameter space.

The only pre-chosen variable is the ISM density. For a choice of ISM density (or range of densities), and for each model (with hydrosim-based corrections applied), an emission radius is generated. Either $A_{\star}$ or $n$ is noted depending on where $r_{\rm emit}$ is at 11 hours, and the CSM structure for that model at core-collapse. In this way, for each choice of ISM density, we can generate synthetic observed distributions of $A_{\star}$ and $n$ using plausible progenitor models and realistic distributions of redshifts and burst energetics. 

We list the results of this exercise in the upper part of Table \ref{tab:results}. ISM densities from 10$^{-3}$ to 10$^{7}$\,cm$^{-3}$ are trialled. This covers a wide range of environments, from galactic halos to the densest molecular clouds \citep{2001RvMP...73.1031F}. To compare to the observed sample, we compare the predicted ratio of ISM to wind-like bursts (observationally 1.55$\pm$0.37, see Section \ref{sec:obsv}), and how well the distribution of $A_{\star}$, $n$ and $r_{\rm emit}$ compare with observations. The KS test p-values from these comparisons are used to gauge how well the distributions match relative to other densities, rather than evaluate a true goodness of fit. This is because we are using fixed ISM densities, with large logarithmic gaps between them, artificially two-step distributions (stalled and ISM densities) in our predictions, which will never agree well with observations. Nevertheless, as the Table \ref{tab:results} shows, it is difficult to get agreement with every observable simultaneously.

\begin{figure*}
\centering
\includegraphics[width=0.99\textwidth]{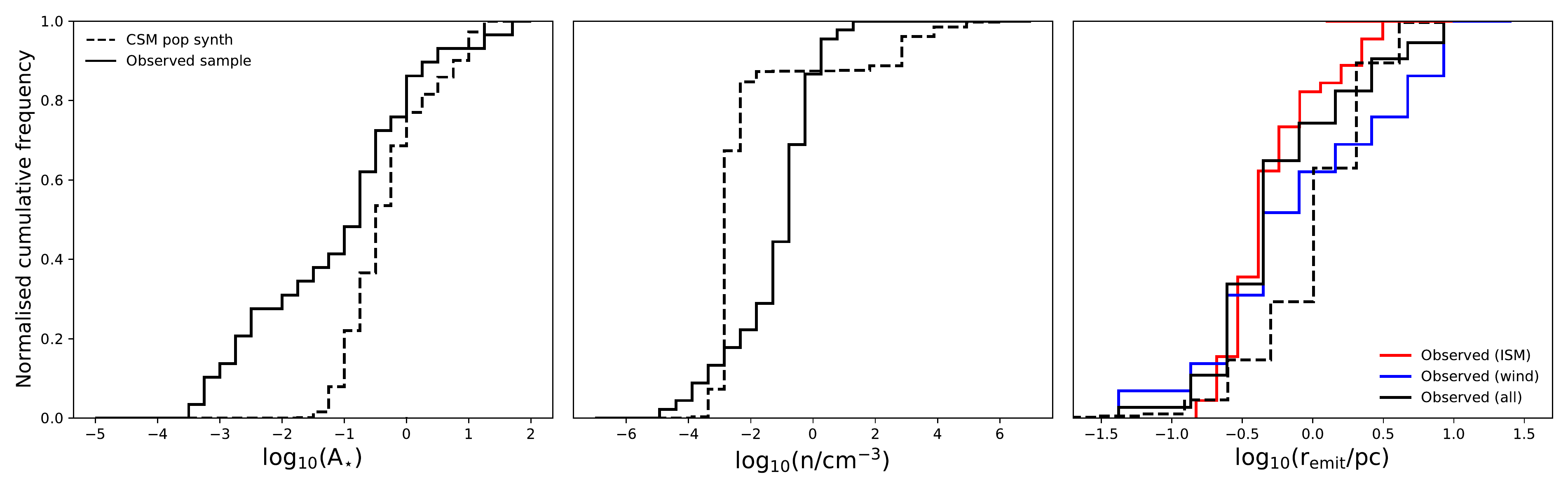}
\caption{The CSM population synthesis distributions for \astar\ , $n$ (including stalled wind and ISM contributions) and r$_{\rm emit}$ versus observations, for a Gaussian distribution in log ISM density, centered at log$_{10}($n/cm$^{-3})=3$, the best matching single density with our standard assumptions (see Table \ref{tab:results}). }
\label{fig:bestmatch}
\end{figure*}

We now discuss which is the best single ISM density, based on the statistical comparison of the three distributions ($n$, \astar\, $r_{\rm emit}$), and the ISM/wind ratio. To evaluate the best-fit log density, for each of $n$, \astar\, and $r_{\rm emit}$, we calculate the weighted mean density across the range of ISM densities trialled, weighted by the p-value. The resultant weighted mean densities are log$_{10}$(n/cm$^{-3}$)=4.1, --0.9, and 6.6 for $n$, \astar\, and $r_{\rm emit}$ respectively. The ISM/wind ratio is the only parameter to have a single, clear maximum in probability across this density range, occurring at log$_{10}$(n/cm$^{-3}$)$\sim$3. As this value is also the mean of the three weighted mean densities from $n$, \astar\, $r_{\rm emit}$, we conclude that $\sim$10$^{3}$\,cm$^{-3}$ is overall the best matching single density. The least-well fit parameter is $n$, because at a single ISM density, we only produce two $n$ values (the stalled wind and the ISM itself). To investigate if a range of densities improves this, we also try a Gaussian distribution of log densities, centred at 3. This yields an ISM/wind ratio of 1.08, with \astar\  and r$_{\rm wind}$ p-values of --1.37 and --4.67, similar to the $n=1000$\,cm$^{-3}$ result, while the $n$ p-value is improved as expected, giving $p = -1.77$. The results obtained for this Gaussian distribution of ISM densities are shown in Figure \ref{fig:bestmatch}.

In Figure \ref{fig:remit}, we show normalised histograms of the emission radii for both ISM and wind observations, plus ISM and wind-like population synthesis predictions (at $n=1000$\,cm$^{-3}$). In the predictions, while there is some overlap, emission in ISM environments is strongly biased to higher radii than wind-like environments. However, in the observations, wind-like environments occur out to higher radii than ISM environments. This implies the existence of substantial variety in the wind strengths of the progenitors, ISM densities, or both. Since the maximum of the synthetic \astar\ distribution is similar to that observed, a population occurring in lower-density environments seems the best explanation for the wind-like environments at large radii. However, when averaged over the population, we find that high density environments are required.

Supernova remnants in the Milky Way favour environments no denser than $n=10$\,cm$^{-3}$, and typically lower in the Magellanic clouds \citep{2022arXiv220505103A}, so unless core-collapse GRBs favour denser environments than regular core-collapse supernovae, it is likely that $r_{\rm wind}$ and/or A$_{\star}$ have been overestimated by our population synthesis. The finding that high densities are needed to match the $r_{\rm wind}$ values inferred from GRB afterglows, given otherwise reasonable assumptions regarding the progenitors and mass loss rates, has previously been noted \citep[e.g.][]{2006MNRAS.367..186E,2006A&A...460..105V}. We have shown this problem to persist at a population level, and now discuss possible explanations for the discrepancy.

\begin{table}
\centering 
\caption{Synthetic distributions for \astar\, constant densities (including both the stalled wind region and ISM) and $r_{\rm emit}$ are generated for the following choices of ISM density. The results are compared against the following observables. First, the number of $\sigma$ away from the observed ratio of constant density to wind-like bursts (1.55$\pm$0.37, Poisson uncertainties). Secondly, the distributions of \astar\, $n$ and $r_{\rm emit}$ through KS tests. The log of the KS test p-value is listed. The null hypothesis that they are drawn from the same distribution is rejected for log$_{10}$(p)$<-1.3$ ($p<0.05$). In the final row, we list results for a Gaussian distribution in log$_{10}$(n), centered on 3. Minima in the distributions as a function of density are shown in bold.} 
\label{tab:results}
\begin{tabular}{llllll}
\hline %
log$_{10}$(n/cm$^{3}$) & Ratio & $\sigma_{\rm{ratio}}$ & $n$ p-val & $r_{\rm emit}$ p-val & \astar\ p-val \\									
\hline %
--3	&	0.04	&	4.10	&	--15.7	&	--5.89	&	--1.14	\\
--2	&	0.04	&	4.10	&	--13.4	&	--6.29	&	\bf{--1.13}	\\
--1	&	0.04	&	4.10	&	--8.38	&	--5.78	&	\bf{--1.13}	\\
0	&	0.05	&	4.07	&	--3.39	&	--5.55	&	--1.15	\\
1	&	0.12	&	3.86	&	--6.50	&	--5.32	&	--1.39	\\
2	&	0.43	&	3.05	&	--9.96	&	--5.65	&	--1.53	\\
3	&	\bf{1.12}	&	\bf{1.18}	&	--8.49	&	--5.32	&	--2.08	\\
4	&	2.69	&	3.08	&	--6.64	&	--5.58	&	--3.80	\\
5	&	6.43	&	13.2	&	\bf{--2.76}	&	--5.18	&	--6.46	\\
6	&	11.7	&	27.5	&	--5.03	&	--3.94	&	--9.61	\\
7	&	18.9	&	47.0	&	--12.2	&	\bf{--3.28}	&	--10.7	\\
\hline
Gauss., $\mu=3$	&	1.08	&	1.27	&	--1.77	&	--4.67	&	--1.37	\\
\hline %
\end{tabular}
\end{table}

\section{Discussion}\label{sec:discuss}
We now discuss possible explanations for the discrepancy with observation, and  whether we can explain the observed CSM variety around long GRB progenitors.

\subsection{Radiative cooling}
In all 16 simulations used for the analytic corrections (Fig. \ref{fig:hydrosims}), our cooling was fixed at the initial ZAMS metallicity and not varied. The cooling rate in a Wolf-Rayet wind can be significantly increased, by a factor of 10-100, due to the enhanced abundance of elements heavier than hydrogen, but this does not translate into a large impact in, for example, $r_{\rm wind}$. \citet{2002A&A...394..901M} demonstrate that, with respect to Solar metallicity outflows with the same mass loss rate and velocity, winds from Wolf-Rayet stars produce termination shock radii that are $\sim$1.5 times lower (with significant scatter depending on the exact composition of the wind). However, $\dot{M}$ and $V_{\rm wind}$ can be 10-100 times higher than on the main sequence, which can more than compensate and in some cases dominate over the cooling differences (see equations in Section \ref{sec:analytic}). Finally, we can see in Fig. \ref{fig:verification} a simulation which matches the assumptions of the analytic model in every regard, except for the inclusion of cooling. We find $r_{\rm wind}=4.5$\,pc (with cooling) versus 4.9\,pc (without). Therefore, we conclude that the precise nature of the cooling implemented, or indeed even if it is included or not, is not a dominant factor in driving the CSM structure - particularly as the abundance variation during model evolution applies primarily only to late-stage Wolf-Rayet outflows (if such a phase occurs). 

\subsection{Binarity}
In our simulations, we only included the wind from the progenitor star, without including the companion. The orbital separations are much less than the parsec-scale bubbles, so we can consider the winds from the binary as a single outflow at termination shock distances, even if complex interactions occur on smaller spatial scales \citep[such as colliding winds,][]{2010ApJ...716L.223G}. More generally, secondary star winds are typically weaker (secondaries are less massive), and even in the twin case, $\dot{M}$ is only increased by a factor of two. Since $r_\mathrm{wind} \propto \sqrt{\dot{M}}$, including the secondary wind would increase $r_\mathrm{wind}$ by at most a factor of 1.4 (Equation \ref{eq:rw}). Another effect of binarity is that the secondary wind, post primary-SN, blows not into the ISM but the wind bubble of the combined primary plus secondary wind up to that point. However, we note that this is only relevant if the system remains bound. For unbound companion velocities of 50\,kms$^{-1}$, and secondary lifetimes post-primary of 10$^{7}$\,yr (Fig. \ref{fig:DTD}), the unbound companion will be $\sim$500\,pc away by the time it undergoes core-collapse. Given typical stalled wind-ISM boundary radii of $\sim$1\,pc (for a 1\,cm$^{-3}$ ISM, e.g. Figure \ref{fig:Rwind}), this is far outside the sphere of influence of the original system. Since most binary systems are unbound upon primary supernova \citep[e.g.][]{2011MNRAS.414.3501E,2019A&A...624A..66R,2022MNRAS.tmp.1065C}, the assumption of similar ISM properties for primaries and secondaries should generally hold.

\begin{figure}
\centering
\includegraphics[width=0.48\textwidth]{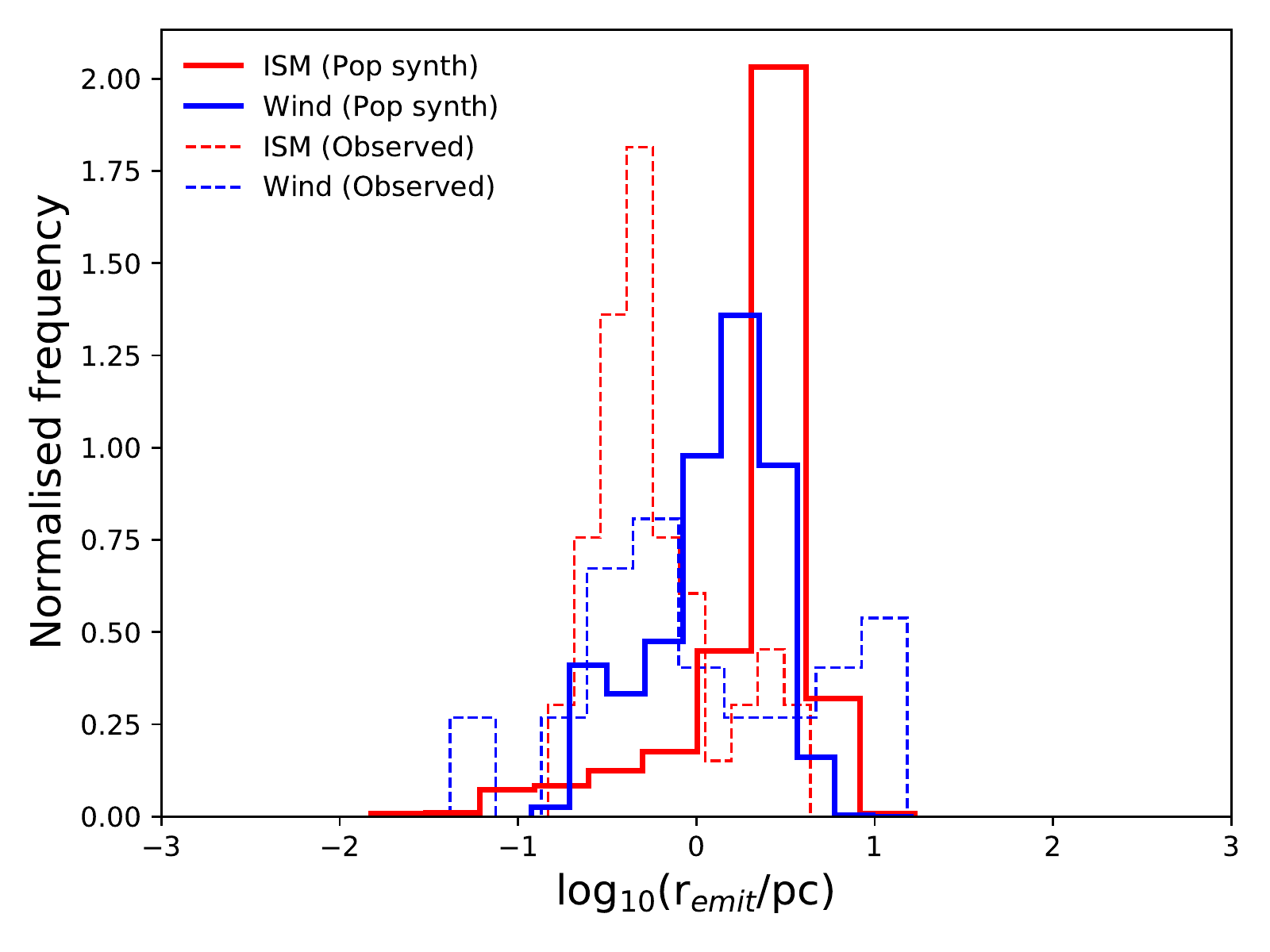}
\caption{Histograms of the emission radii from both the population synthesis predictions of the best-fit ISM density distribution (solid lines, ISM in red, wind in blue) and observations (dashed lines). The observed sample is more mixed, with wind-like environment found at higher radii than constant density environments. This implies a broader range of progenitor wind strengths and/or densities than simulated.}
\label{fig:remit}
\end{figure}

\subsection{Interstellar magnetic fields}
Interstellar magnetic fields are not included in the simulations, but also have an effect on the expansion of wind blown bubbles, confining them in the direction perpendicular to the field lines. Although dependent on the field and wind strengths, \citet{2015A&A...584A..49V} show that the compression along this axis (and resultant decrease in $r_\mathrm{wind}$) is typically a factor of a few for magnetic fields of ${\sim}10-20\,{\mu}$G. These field strengths can be found in molecular clouds or close to Galactic centre, but typical Galactic magnetic fields are ${\sim}5\,{\mu}$G \citep{2011NewAR..55...91V}.

\subsection{Wind bubble asymmetry}
Motion through the ISM also has an effect on shock structure, but this is primarily on the outer shock structure, rather than on the inner termination shock at $r_\mathrm{wind}$ \citep[e.g.,][]{2006A&A...460..105V,2020MNRAS.493.3548M}. Furthermore, the compression/rarefaction of the wind bubble is most significant along the direction-of-travel. If the stars with the fastest velocities have been kicked by the supernova of a companion \citep{2007A&A...465L..29C,2011MNRAS.414.3501E}, we might expect them to have been kicked preferentially in the orbital plane, which is typically aligned with the orbital axes. In this case, the jets would tend to be perpendicular to the direction of motion, a direction less affected (on average) by the presence of a bow shock structure. In superbubble environments with hot and tenuous gas, bow shocks may not form at all \citep{2002A&A...383..999H}. Another factor is progenitor rotation, which may lead to asymmetric winds, with the mass loss rate decreased in the polar direction. Since this is the axis along which jets are expected to be launched, if asymmetric mass loss persists until close to core-collapse, the jet may see a termination shock radius a factor of few closer than for spherically symmetric mass loss \citep{1996ApJ...459..671I,2007MNRAS.377L..29E}.

\subsection{Eruptive and irregular mass loss}
Another consideration is whether eruptive very-late stage mass loss plays a role in shaping the observed afterglow evolution. Such eruptive episodes are thought to produce the dense CSM environments responsible for type IIn supernovae \citep{2014ARA&A..52..487S}. The BPASS models used evolve up to the end of core carbon burning. The neon, oxygen and subsequent core burning stages together last only a few years. Assuming eruptive mass loss with strong winds of 3000\,kms$^{-1}$ during this time, the furthest any shells ejected post-model is $\sim$0.05\,pc \citep[see also][]{2021ApJ...907...99S}. Given that the standardised (cosmological) rest-frame time of 11 hours corresponds to typical jet propagation times and distances post-burst measured in years and parsecs, any impact on the afterglow evolution will be extremely early (in our reference frame) and almost never driving the observed afterglow behaviour. Similarly, late-stage common envelope evolution could eject a large amount of material close to core-collapse. In supernovae with relatively slow moving ejecta, this can influence the lightcurve evolution, for example by producing a bump as the ejecta reaches the previously-ejected common envelope material \citep[e.g.][]{2021MNRAS.504L..51S}. In GRBs however, the jet passage across any region of envelope ejecta will be rapid and unlikely to shape the afterglow evolution in a meaningful way.
Finally, it is possible that the stellar wind deviates from an $r^{-2}$ profile at late times, since the rapid evolution of a star could change the stellar wind on a timescale faster than an equilibrium profile can established. This is particularly true for rapidly rotating stars near critical rotation \citep{2008A&A...478..769V}. However, such rapid variation is again expected extremely late in the progenitor's life, and will therefore affect the CSM at very low radii, such that the jet will likely traverse these regions before observations are obtained in our reference frame.

\subsection{Correlation between isotropic-equivalent energy and progenitor properties}\label{sec:testrwindeiso}
It is possible that our random assignment of E$_{\rm iso}$ to each model is missing some correlation between wind properties and burst energy. To investigate, we try drawing initial E$_{\rm iso}$ values following E$_{\rm iso} \propto r_{\rm wind}$ and E$_{\rm iso} \propto 1/r_{\rm wind}$. The first represents the case where more massive stars, with stronger winds, produce stronger GRBs. The second case assumes that stronger winds correlate with weaker bursts (e.g., less angular momentum left in progenitor). In each case, the correlation is parameterised as a log-lin relation (log in E$_{\rm iso}$ and lin in r$_{\rm wind}$), with the maximum E$_{\rm iso}$ in Section \ref{sec:obsv} corresponding to the the maximum or minimum r$_{\rm wind}$, and vice versa. As in Section \ref{sec:results}, we calculate the p-values and ISM/wind ratio at each of the log$_{10}$(n) values and determine the p-value weighted mean log density, across this range, in each case. Under the assumption that  E$_{\rm iso} \propto r_{\rm wind}$, we find weighted mean log densities of --1.1, 3.4 and 6.9 for \astar, $n$ and r$_{\rm emit}$ respectively. The ISM/wind ratio favours ${\rm n}=100$\,cm$^{-3}$, with ISM/wind $= 1.29$ at this density. The quality of the fits overall is no better than the default random assignment of E$_{\rm iso}$, and is worse for r$_{\rm emit}$, with typical p-values around log(p)$\sim$--9. For E$_{\rm iso} \propto 1/r_{\rm wind}$, we find weighted mean log densities of --0.8, 4.5 and 6.3 respectively, with the best ISM/wind $=1.11$ ratio again at ${\rm n}=1000$\,cm$^{-3}$. Overall, there is no clear indication that the progenitor wind properties are correlated or anti-correlated with E$_{\rm iso}$, but on the other hand, this demonstrates that our methodology is robust against assumptions of this nature. We note, however, that we cannot rule out some other correlation with progenitor properties playing a role.

\subsection{Mass loss rates and terminal wind velocities}
One possibility is that the models selected are not truly representative of the collapsar GRB population. Since we need to decrease the termination shock radius to match observations, and mass loss rates and wind speeds decrease at lower metallicity, we check whether our models are not sufficiently biased to low metallicity by selecting only models at $Z/Z_{\odot}<0.3$. At ${\rm n}=10$\,cm$^{-3}$, we find an ISM/wind ratio of 0.11, with \astar, n, and $r_{\rm{emit}}$ log p-values of --1.38, --5.57 and --7.8 - similar to the results when all metallicities are considered, as in Table \ref{tab:results}. The reason for this can be seen in Figs. \ref{fig:Rwind}, \ref{fig:Astar} and \ref{fig:DTD}. Although the lower metallicity models, with include QHE models, have weaker winds, they on average have longer lifetimes. These effects balance, leading to comparable terminal shock radii at the time of core-collapse. Therefore, selecting subsets of our models by metallicity cannot resolve the observed discrepancy. 

Alternatively, a reduction in mass loss rates and/or wind speeds more generally may provide a solution. For example, compared to the adopted OB star mass loss rates of \cite{2001A&A...369..574V}, \cite{2021A&A...648A..36B,2022arXiv220308218B} find mass loss rates that are a factor of three lower. There is substantial uncertainty in Wolf-Rayet mass loss rates and wind speeds, and the exact point at which Wolf-Rayet behaviour ceases \citep{2020MNRAS.491.4406S,2020MNRAS.499..873S}. Furthermore, in Fig. \ref{fig:bestmatch}, we can see that the population synthesis fails to predict the lowest $\sim$third of observed \astar\ values. Using the analytic approximations in Section \ref{sec:analytic}, for constant wind speeds, $r_{\rm wind} \propto \sqrt{\dot{M_{w}}}$ and vice versa for constant mass loss rates. As an example, for a lower ${\rm n}=10$\,cm$^{-3}$, artificially reducing $r_{w}$ and $r_{sw}$ by a factor $\sim$three corresponds to late stage mass loss or wind speeds being reduced by a factor of ten. Applying these corrections, we find an ISM/wind ratio of 1.38 ($\sigma_{\rm ratio} = 0.48$), and log p-values of --0.73 4.28 and --5.57 for \astar, n, and $r_{\rm{emit}}$, overall in better agreement than the default ${\rm n}=10$\,cm$^{-3}$ results.

\subsection{Summary}
All of the considerations above could play a role in bridging the gap between observations and modelling, assuming that long GRB progenitors actually explode in lower ISM densities than inferred from our modelling. ISM densities of $100 < {\rm n}/$\,cm$^{-3} < 1,000,000$ are plausible if core-collapse GRBs occur close to their birth sites in dense molecular clouds \citep{2001RvMP...73.1031F}. However, it is expected that the densest cores do not last more than a few Myr \citep[e.g.][]{2005ApJ...618..344V}, comparable or shorter than the lifetimes of the progenitors (see Fig. \ref{fig:DTD}). Such dense molecular cloud environments are also expected to be extremely dusty, with high neutral hydrogen column densities, and may also be conducive to producing VHE emission in GRBs \citep[e.g.][]{2022MNRAS.tmp.1037R}. Given the measured N$_{H}$ values in GRBs \citep{Evans09}, prevalence of optically-detected GRBs and rarity of VHE-detected events, qualitatively it appears as though the high densities inferred from our modelling are instead indicative of flaws in the model. Furthermore, our observational sample is biased against dark GRBs, compared with the overall long GRB population \citep[dark fraction $25-40\%$,][]{2009ApJS..185..526F,2011A&A...526A..30G}, due to the requirement of well-detected optical lightcurves as a pre-requisite. If we favour high density environments for the optically bright GRBs, this implies even higher densities for the overall population, if dark bursts are taken into account. Overall, there is no one factor which can clearly make up the difference between observation and theory. Even for a lower ISM density and reduced wind strengths (as in our ${\rm n}=10$\,cm$^{-3}$ example above), substantial variation in the termination shock radius is required to reproduce the observed range of wind and ISM environments at different emission radii (Figure \ref{fig:remit}).

\section{Conclusions}\label{sec:conclusion}
In this paper, we have performed a circumstellar medium population synthesis for long GRB progenitors. A semi-analytic wind-blown bubble model was applied to BPASS binary stellar evolution models, to derive termination shock radii where the stellar wind transitions into a flat, shocked wind profile. Through comparison with a grid of hydrodynamical simulations, we derive the outer shocked wind-ISM interface radius for each analytic result, and correct the termination shock radius based on trends in the difference between values derived through analytic and hydrodynamical methods. The results for $r_\mathrm{wind}$ and \astar\ were then compared to the environments, densities and emission radii inferred from the largest long GRB afterglow dataset yet compiled, evaluated at a rest-frame time of 11 hours. We find that, under standard assumptions for the mass loss rates and terminal wind velocities of likely progenitor stars, high ISM densities of ${\rm n} \sim 1000$\,cm$^{-3}$ best reproduce the observed distributions of $r_\mathrm{wind}$ and \astar\ while also reproducing the observed ratio of ISM to wind-like environments. Given the lack of dark GRBs in our sample, the mean ISM density for the overall population is likely even higher. However, if long GRBs occur in more typical ISM environments (${\rm n}\sim 1 $\,cm$^{-3}$), our predictions for the termination shock radii must be reduced, confirming the findings of individual stellar model studies at a population level. We find that selecting subsets of models by metallicity or the type of progenitor system cannot resolve the discrepancy. Instead, a range of factors from reduced mass loss rates and wind speeds to magnetic field confinement could plausibly be contributing to this persistent gap between GRB observation and theory.

\section*{Acknowledgements}
AAC is supported by the Radboud Excellence Initiative. BPG and MN are supported by the European Research Council (ERC) under the European Union’s Horizon 2020 research and innovation programme (grant agreement No.~948381). DAK acknowledges support from Spanish National Research Project RTI2018-098104-J-I00 (GRBPhot). AJvM is supported by the  ANR-19-CE31-0014GAMALO project. AJL has received funding from the European Research Council (ERC) under the European Union’s Seventh Framework Programme (FP7-2007-2013) (Grant agreement No. 725246). ERS has been supported by STFC consolidated grant ST/P000495/1. PJG is supported by NRF SARChI Grant 111692. We gratefully acknowledge the use of {\sc gotohead}, the computing cluster of the Gravitational-wave Optical Transient Observer (GOTO), as well as support from Joe Lyman and Krzysztof Ulaczyk.

This work made use of v2.2.1 of the Binary Population and Spectral Synthesis (BPASS) models as described in \cite{2017PASA...34...58E} and \cite{2018MNRAS.479...75S}. This work made use of data supplied by the UK Swift Science Data Centre at the University of Leicester. This work made use of the The PLUTO Code for Astrophysical GasDynamics \citep{2007ApJS..170..228M}. This work has made use of {\sc ipython} \citep{2007CSE.....9c..21P}, {\sc numpy} \citep{2020arXiv200610256H}, {\sc scipy} \citep{2020NatMe..17..261V}; {\sc matplotlib} \citep{2007CSE.....9...90H} and {\sc astropy},\footnote{\url{https://www.astropy.org}} a community-developed core Python package for Astronomy \citep{astropy:2013, astropy:2018}. Finally, we thank the referee for their careful consideration of the manuscript.

\section*{Data Availability}
The BPASS binary evolution models used in this work are available via the project website\footnote{\url{https://bpass.auckland.ac.nz/}}. The post-processed models \citep[as described in][]{2020MNRAS.491.3479C} and raw data for the CSM population synthesis predictions presented in this paper are available upon reasonable request. The results from the GRB afterglow fitting are available in Appendix~\ref{app:fits}, along with example plots showing fits in wind-like and ISM-like media. Equivalent plots for every GRB in the sample are available as supplementary material on the journal website.



\bibliographystyle{mnras}
\bibliography{grbwinds} 



\newpage\clearpage

\onecolumn

\appendix
\section{Afterglow fit results}

\renewcommand*{\arraystretch}{1.3}

\begin{center}
\begin{longtable}{|c|c|c|c|c|c|c|c|c|}
\caption{MCMC posterior median values and $1\sigma$ confidence intervals for our sample of GRB afterglows. Our observed sample constitutes 75 GRBs with unique solutions for their environment types. A further 13 GRBs display acceptable fits to either environment. These are tabulated here but not used in our analysis. ISM fits have density units of cm$^{-3}$, while wind fits have $5\times10^{11}$\,g\,cm$^{-1}$.}
\label{app:fits}\\

\hline\hline
 & & & & & & & & Radio \\
GRB & Environment & log(E$_{\rm k,iso}$/erg) & p & log($\epsilon_b$) & log(n/cm$^{-3}$) & log(A$_*$/$5\times10^{11}$\,g\,cm$^{-1}$) & $\theta_o$ (rad) & data \\
\hline\hline
\endfirsthead

\hline\hline
 & & & & & & & & Radio \\
GRB & Environment & log(E$_{\rm k,iso}$/erg) & p & log($\epsilon_b$) & log(n/cm$^{-3}$) & log(A$_*$/$5\times10^{11}$\,g\,cm$^{-1}$) & $\theta_o$ (rad) & data \\
\hline\hline
\endhead

\hline\hline
\endfoot

050315 & ism & 55.44$_{-0.37}^{+0.30}$ & 2.01$_{-0.002}^{+0.004}$ & -4.05$_{-0.60}^{+0.69}$ & -3.34$_{-0.39}^{+0.45}$ & ----- & 0.02$_{-0.003}^{+0.006}$ & [1] \\
050318 & ism & 53.08$_{-0.36}^{+0.45}$ & 2.19$_{-0.09}^{+0.05}$ & -3.94$_{-0.68}^{+1.17}$ & 0.12$_{-1.64}^{+1.30}$ & ----- & 0.04$_{-0.01}^{+0.03}$ & \\
050319 & ism & 54.46$_{-0.16}^{+0.11}$ & 2.01$_{-0.003}^{+0.010}$ & -4.30$_{-0.52}^{+0.69}$ & -0.17$_{-1.07}^{+0.81}$ & ----- & 0.08$_{-0.02}^{+0.02}$ & \\
050416A & ism & 54.11$_{-0.66}^{+0.55}$ & 2.01$_{-0.002}^{+0.004}$ & -3.74$_{-0.91}^{+1.12}$ & -4.45$_{-0.38}^{+0.53}$ & ----- & 0.28$_{-0.15}^{+0.15}$ & [1] \\
050505 & wind & 53.68$_{-0.71}^{+0.07}$ & 2.19$_{-0.17}^{+0.06}$ & -1.15$_{-0.27}^{+0.37}$ & ----- & 1.77$_{-0.90}^{+0.18}$ & 0.30$_{-0.03}^{+0.11}$ & \\
050525A & wind & 52.72$_{-0.49}^{+0.16}$ & 2.63$_{-0.53}^{+0.03}$ & -4.79$_{-0.17}^{+3.98}$ & ----- & -0.17$_{-0.18}^{+0.44}$ & 0.43$_{-0.20}^{+0.05}$ & [2,3] \\
050801 & ism & 51.49$_{-0.17}^{+0.23}$ & 2.10$_{-0.05}^{+0.53}$ & -0.92$_{-1.14}^{+0.32}$ & 1.07$_{-1.71}^{+0.63}$ & ----- & 0.38$_{-0.13}^{+0.09}$ & \\
050802 & wind & 54.16$_{-1.11}^{+1.27}$ & 2.26$_{-0.15}^{+0.02}$ & -2.34$_{-1.04}^{+1.12}$ & ----- & -3.17$_{-0.08}^{+4.08}$ & 0.27$_{-0.15}^{+0.16}$ & \\
050820A & ism & 54.31$_{-0.10}^{+0.10}$ & 2.29$_{-0.01}^{+0.01}$ & -4.91$_{-0.07}^{+0.15}$ & -0.48$_{-0.23}^{+0.21}$ & ----- & 0.13$_{-0.01}^{+0.01}$ & [2,4] \\
050824 & ism & 53.17$_{-0.38}^{+0.37}$ & 2.00$_{-0.002}^{+0.005}$ & -1.87$_{-1.15}^{+0.95}$ & -1.54$_{-1.42}^{+1.70}$ & ----- & 0.36$_{-0.13}^{+0.10}$ & [2,5] \\
050908 & ism & 52.20$_{-0.40}^{+0.60}$ & 2.82$_{-0.20}^{+0.11}$ & -1.46$_{-0.88}^{+0.67}$ & -4.06$_{-0.58}^{+0.55}$ & ----- & 0.27$_{-0.16}^{+0.16}$ & \\
050922C & ism & 53.21$_{-0.3}^{+0.41}$ & 2.75$_{-0.02}^{+0.02}$ & -4.53$_{-0.35}^{+0.84}$ & -0.64$_{-1.51}^{+0.83}$ & ----- & 0.10$_{-0.04}^{+0.04}$ & [2] \\
060124 & wind & 53.51$_{-0.01}^{+0.01}$ & 2.33$_{-0.01}^{+0.01}$ & -0.51$_{-0.02}^{+0.01}$ & ----- & -2.70$_{-0.01}^{+0.02}$ & 0.29$_{-0.13}^{+0.15}$ & \\
060418 & wind & 54.01$_{-0.91}^{+0.56}$ & 2.08$_{-0.03}^{+0.04}$ & -4.16$_{-0.62}^{+2.03}$ & ----- & -0.39$_{-0.41}^{+0.35}$ & 0.19$_{-0.05}^{+0.07}$ & [2,6] \\
060512 & ism & 50.86$_{-0.14}^{+0.39}$ & 2.13$_{-0.09}^{+0.12}$ & -1.09$_{-0.61}^{+0.43}$ & -1.02$_{-0.65}^{+0.90}$ & ----- & 0.41$_{-0.09}^{+0.06}$ & \\
060526 & ism & 52.83$_{-0.10}^{+0.08}$ & 2.26$_{-0.01}^{+0.01}$ & -2.88$_{-1.00}^{+1.08}$ & -0.10$_{-1.27}^{+1.43}$ & ----- & 0.10$_{-0.04}^{+0.05}$ & \\
060607A & wind & 53.49$_{-0.53}^{+0.76}$ & 2.96$_{-0.07}^{+0.03}$ & -0.85$_{-1.53}^{+0.19}$ & ----- & -2.40$_{-0.16}^{+0.30}$ & 0.05$_{-0.02}^{+0.02}$ & \\
060714 & ism & 53.32$_{-0.22}^{+0.31}$ & 2.22$_{-0.0001}^{+0.0002}$ & -4.37$_{-0.45}^{+0.75}$ & -0.51$_{-0.83}^{+1.04}$ & ----- & 0.07$_{-0.02}^{+0.03}$ & \\
060729 & wind & 53.86$_{-0.01}^{+0.01}$ & 2.01$_{-0.001}^{+0.001}$ & -4.98$_{-0.01}^{+0.02}$ & ----- & 1.35$_{-0.01}^{+0.00}$ & 0.50$_{-0.001}^{+0.001}$ & \\
060904B & ism & 51.94$_{-0.56}^{+1.09}$ & 2.68$_{-0.13}^{+0.05}$ & -2.41$_{-1.38}^{+1.09}$ & -3.01$_{-1.31}^{+2.61}$ & ----- & 0.30$_{-0.15}^{+0.14}$ & \\
061007 & wind & 52.46$_{-0.10}^{+0.13}$ & 2.38$_{-0.11}^{+0.07}$ & -3.01$_{-1.21}^{+1.13}$ & ----- & -0.72$_{-0.82}^{+1.02}$ & 0.34$_{-0.10}^{+0.11}$ & \\
061121 & wind & 55.76$_{-0.88}^{+0.18}$ & 2.00$_{-0.002}^{+0.18}$ & -1.02$_{-0.81}^{+0.37}$ & ----- & -2.92$_{-2.06}^{+2.36}$ & 0.24$_{-0.13}^{+0.18}$ & [7] \\
061126 & ism & 53.18$_{-0.12}^{+0.24}$ & 2.04$_{-0.02}^{+0.03}$ & -4.01$_{-0.72}^{+1.01}$ & -0.05$_{-1.54}^{+1.14}$ & ----- & 0.10$_{-0.04}^{+0.04}$ & \\
070125 & ism & 54.01$_{-1.27}^{+0.01}$ & 2.89$_{-0.01}^{+0.01}$ & -4.99$_{-0.01}^{+3.28}$ & -0.18$_{-0.73}^{+0.03}$ & ----- & 0.25$_{-0.01}^{+0.12}$ & [8] \\
071003 & wind & 52.90$_{-0.05}^{+0.06}$ & 2.11$_{-0.02}^{+0.03}$ & -0.58$_{-0.17}^{+0.06}$ & ----- & -0.97$_{-0.09}^{+0.38}$ & 0.18$_{-0.01}^{+0.02}$ & [9] \\
071010A & ism & 52.14$_{-0.04}^{+0.04}$ & 2.18$_{-0.02}^{+0.02}$ & -2.32$_{-0.35}^{+0.62}$ & 1.29$_{-0.93}^{+0.52}$ & ----- & 0.19$_{-0.04}^{+0.03}$ & \\
----- & wind & 52.47$_{-0.08}^{+0.05}$ & 2.15$_{-0.02}^{+0.04}$ & -3.77$_{-0.27}^{+0.95}$ & ----- & 1.19$_{-0.41}^{+0.18}$ & 0.46$_{-0.06}^{+0.03}$ & \\
071031 & ism & 52.50$_{-0.26}^{+1.05}$ & 2.54$_{-0.15}^{+0.07}$ & -3.60$_{-0.92}^{+1.21}$ & -0.97$_{-1.40}^{+1.67}$ & ----- & 0.28$_{-0.15}^{+0.15}$ & \\
071112C & ism & 52.21$_{-0.59}^{+1.15}$ & 2.32$_{-0.04}^{+0.03}$ & -3.05$_{-1.35}^{+1.58}$ & -2.60$_{-1.58}^{+2.24}$ & ----- & 0.30$_{-0.15}^{+0.14}$ & \\
080210 & wind & 52.83$_{-0.19}^{+0.44}$ & 2.15$_{-0.10}^{+0.18}$ & -2.50$_{-1.61}^{+1.38}$ & ----- & -0.57$_{-1.07}^{+1.15}$ & 0.34$_{-0.12}^{+0.11}$ & \\
080310 & ism & 52.80$_{-0.08}^{+0.07}$ & 2.40$_{-0.01}^{+0.01}$ & -3.72$_{-0.73}^{+0.88}$ & 0.53$_{-1.24}^{+1.05}$ & ----- & 0.15$_{-0.04}^{+0.05}$ & \\
080319B & ism & 53.42$_{-0.21}^{+0.20}$ & 2.66$_{-0.004}^{+0.004}$ & -4.80$_{-0.15}^{+0.34}$ & -0.84$_{-0.48}^{+0.38}$ & ----- & 0.26$_{-0.05}^{+0.05}$ & [10-13] \\
080413B & ism & 53.42$_{-0.05}^{+0.03}$ & 2.01$_{-0.001}^{+0.001}$ & -3.37$_{-0.83}^{+1.07}$ & 0.34$_{-1.60}^{+1.24}$ & ----- & 0.15$_{-0.05}^{+0.06}$ & \\
080603A & wind & 53.45$_{-1.03}^{+0.19}$ & 2.03$_{-0.01}^{+0.19}$ & -0.64$_{-0.30}^{+0.11}$ & ----- & -1.49$_{-0.15}^{+0.32}$ & 0.12$_{-0.01}^{+0.16}$ & [14] \\
----- & ism & 52.75$_{-0.18}^{+0.16}$ & 2.07$_{-0.02}^{+0.04}$ & -0.59$_{-0.22}^{+0.07}$ & -2.11$_{-0.24}^{+0.41}$ & ----- & 0.06$_{-0.01}^{+0.02}$ & [14] \\
080605 & ism & 53.33$_{-0.22}^{+0.25}$ & 2.03$_{-0.01}^{+0.02}$ & -2.47$_{-1.43}^{+1.32}$ & -1.27$_{-1.97}^{+2.14}$ & ----- & 0.28$_{-0.14}^{+0.14}$ & \\
080710 & wind & 51.94$_{-0.14}^{+0.98}$ & 2.48$_{-0.06}^{+0.05}$ & -1.99$_{-1.29}^{+0.90}$ & ----- & -1.49$_{-0.79}^{+1.34}$ & 0.36$_{-0.14}^{+0.10}$ & \\
----- & ism & 52.36$_{-0.39}^{+0.65}$ & 2.83$_{-0.09}^{+0.04}$ & -3.89$_{-0.77}^{+1.28}$ & -0.27$_{-1.68}^{+1.41}$ & ----- & 0.15$_{-0.07}^{+0.09}$ & \\
080721 & wind & 55.21$_{-0.52}^{+0.50}$ & 2.21$_{-0.01}^{+0.02}$ & -3.53$_{-0.97}^{+1.01}$ & ----- & -1.89$_{-0.76}^{+0.79}$ & 0.27$_{-0.14}^{+0.15}$ & \\
080916C & wind & 54.32$_{-0.74}^{+1.07}$ & 2.32$_{-0.07}^{+0.06}$ & -2.71$_{-1.46}^{+1.27}$ & ----- & -2.28$_{-1.05}^{+1.46}$ & 0.29$_{-0.15}^{+0.13}$ & \\
080928 & wind & 52.73$_{-0.23}^{+0.2}$ & 2.34$_{-0.18}^{+0.66}$ & -4.50$_{-0.44}^{+2.82}$ & ----- & 0.14$_{-0.74}^{+0.67}$ & 0.28$_{-0.14}^{+0.15}$ & \\
----- & ism & 52.33$_{-0.15}^{+0.51}$ & 2.23$_{-0.10}^{+0.12}$ & -2.11$_{-0.86}^{+1.05}$ & 0.64$_{-1.60}^{+1.09}$ & ----- & 0.06$_{-0.02}^{+0.04}$ & \\
081007 & ism & 51.77$_{-0.08}^{+0.13}$ & 2.14$_{-0.05}^{+0.07}$ & -1.23$_{-0.41}^{+0.42}$ & -1.73$_{-0.60}^{+0.55}$ & ----- & 0.13$_{-0.02}^{+0.03}$ & [15] \\
081008 & wind & 52.99$_{-0.18}^{+0.08}$ & 2.16$_{-0.02}^{+0.14}$ & -2.36$_{-1.61}^{+1.12}$ & ----- & -0.97$_{-0.98}^{+1.46}$ & 0.33$_{-0.11}^{+0.11}$ & \\
081029 & wind & 53.59$_{-0.57}^{+0.37}$ & 2.08$_{-0.04}^{+0.22}$ & -3.17$_{-1.14}^{+0.56}$ & ----- & 1.34$_{-0.79}^{+0.55}$ & 0.20$_{-0.03}^{+0.04}$ & \\
----- & ism & 53.25$_{-0.32}^{+0.18}$ & 2.05$_{-0.02}^{+0.13}$ & -1.68$_{-0.77}^{+0.32}$ & 1.55$_{-0.59}^{+0.39}$ & ----- & 0.07$_{-0.01}^{+0.01}$ & \\
090102 & wind & 55.31$_{-3.27}^{+0.50}$ & 2.00$_{-0.003}^{+0.056}$ & -2.61$_{-1.74}^{+1.47}$ & ----- & -1.14$_{-1.61}^{+1.96}$ & 0.06$_{-0.03}^{+0.28}$ & \\
----- & ism & 54.92$_{-0.56}^{+0.32}$ & 2.00$_{-0.002}^{+0.011}$ & -4.67$_{-0.24}^{+0.57}$ & -1.77$_{-0.49}^{+0.68}$ & ----- & 0.02$_{-0.003}^{+0.013}$ & \\
090313 & wind & 53.76$_{-0.33}^{+0.03}$ & 2.03$_{-0.003}^{+0.098}$ & -0.51$_{-1.04}^{+0.01}$ & ----- & -0.72$_{-0.03}^{+0.66}$ & 0.30$_{-0.14}^{+0.14}$ & [16] \\
090323 & wind & 53.68$_{-0.04}^{+0.03}$ & 2.78$_{-0.02}^{+0.01}$ & -4.99$_{-0.01}^{+0.04}$ & ----- & 0.56$_{-0.04}^{+0.03}$ & 0.40$_{-0.06}^{+0.07}$ & [17,18] \\
090328 & wind & 52.94$_{-0.20}^{+0.19}$ & 2.03$_{-0.01}^{+0.63}$ & -3.36$_{-1.62}^{+2.81}$ & ----- & 0.26$_{-2.16}^{+0.74}$ & 0.31$_{-0.18}^{+0.14}$ & [18] \\
090423 & ism & 53.90$_{-0.23}^{+0.24}$ & 2.19$_{-0.05}^{+0.05}$ & -4.86$_{-0.10}^{+0.27}$ & -0.53$_{-0.52}^{+0.39}$ & ----- & 0.10$_{-0.02}^{+0.02}$ & [19] \\
090902B & ism & 52.97$_{-0.03}^{+0.04}$ & 2.28$_{-0.02}^{+0.02}$ & -3.07$_{-0.36}^{+0.29}$ & -1.10$_{-0.42}^{+0.50}$ & ----- & 0.19$_{-0.02}^{+0.03}$ & [18] \\
090926A & ism & 52.95$_{-0.05}^{+0.04}$ & 2.98$_{-0.32}^{+0.01}$ & -1.86$_{-0.69}^{+0.05}$ & -2.62$_{-0.03}^{+2.98}$ & ----- & 0.11$_{-0.003}^{+0.126}$ & \\
----- & wind & 53.46$_{-0.47}^{+0.04}$ & 2.13$_{-0.01}^{+0.39}$ & -4.45$_{-0.42}^{+3.95}$ & ----- & 1.40$_{-3.58}^{+0.31}$ & 0.45$_{-0.15}^{+0.04}$ & \\
091018 & ism & 52.93$_{-0.71}^{+0.17}$ & 2.03$_{-0.01}^{+0.13}$ & -3.20$_{-1.01}^{+0.88}$ & 0.26$_{-1.25}^{+1.34}$ & ----- & 0.09$_{-0.03}^{+0.04}$ & \\
----- & wind & 53.55$_{-0.28}^{+0.27}$ & 2.01$_{-0.001}^{+0.006}$ & -1.53$_{-1.57}^{+0.05}$ & ----- & 1.67$_{-0.02}^{+0.20}$ & 0.21$_{-0.03}^{+0.04}$ & \\
091020 & ism & 53.68$_{-0.09}^{+0.09}$ & 2.57$_{-0.01}^{+0.01}$ & -4.93$_{-0.05}^{+0.11}$ & -0.28$_{-0.21}^{+0.19}$ & ----- & 0.32$_{-0.13}^{+0.13}$ & [20] \\
091024 & ism & 51.49$_{-0.09}^{+0.17}$ & 2.67$_{-0.12}^{+0.23}$ & -2.07$_{-0.73}^{+0.59}$ & 1.26$_{-0.66}^{+0.50}$ & ----- & 0.36$_{-0.08}^{+0.10}$ & \\
----- & wind & 51.86$_{-0.06}^{+0.07}$ & 2.50$_{-0.04}^{+0.05}$ & -2.69$_{-0.41}^{+0.56}$ & ----- & 0.11$_{-0.40}^{+0.33}$ & 0.46$_{-0.05}^{+0.03}$ & \\
091029 & ism & 53.80$_{-0.17}^{+0.17}$ & 2.22$_{-0.01}^{+0.01}$ & -4.68$_{-0.23}^{+0.40}$ & -0.73$_{-0.35}^{+0.50}$ & ----- & 0.06$_{-0.01}^{+0.01}$ & \\
091208B & wind & 52.81$_{-0.23}^{+0.75}$ & 2.10$_{-0.06}^{+0.09}$ & -2.51$_{-1.65}^{+1.34}$ & ----- & -1.49$_{-1.22}^{+1.63}$ & 0.31$_{-0.18}^{+0.13}$ & \\
100219A & wind & 51.90$_{-0.18}^{+2.58}$ & 2.43$_{-0.30}^{+0.17}$ & -1.75$_{-1.48}^{+0.59}$ & ----- & 0.16$_{-1.71}^{+0.38}$ & 0.44$_{-0.21}^{+0.05}$ & \\
100418A & ism & 54.00$_{-0.06}^{+0.05}$ & 2.37$_{-0.003}^{+0.003}$ & -4.74$_{-0.08}^{+0.10}$ & -2.00$_{-0.03}^{+0.02}$ & ----- & 0.49$_{-0.013}^{+0.004}$ & [21] \\
100621A & ism & 52.95$_{-0.14}^{+0.09}$ & 2.01$_{-0.003}^{+0.006}$ & -2.62$_{-0.74}^{+1.13}$ & 0.47$_{-1.70}^{+1.10}$ & ----- & 0.16$_{-0.06}^{+0.06}$ & \\
100906A & ism & 53.01$_{-0.25}^{+0.22}$ & 2.03$_{-0.01}^{+0.03}$ & -2.33$_{-1.08}^{+0.68}$ & -0.51$_{-0.92}^{+1.61}$ & ----- & 0.06$_{-0.01}^{+0.03}$ & [22,23] \\
101219B & ism & 51.32$_{-0.04}^{+0.04}$ & 2.26$_{-0.01}^{+0.01}$ & -1.75$_{-0.14}^{+0.15}$ & -0.58$_{-0.15}^{+0.14}$ & ----- & 0.45$_{-0.05}^{+0.04}$ & [24] \\
110205A & wind & 52.88$_{-0.03}^{+0.03}$ & 2.44$_{-0.01}^{+0.01}$ & -4.20$_{-0.09}^{+0.12}$ & ----- & 0.07$_{-0.07}^{+0.06}$ & 0.24$_{-0.01}^{+0.01}$ & [25,26] \\
110213A & wind & 53.71$_{-0.51}^{+0.68}$ & 2.56$_{-0.03}^{+0.02}$ & -3.97$_{-0.69}^{+0.62}$ & ----- & -1.15$_{-0.09}^{+0.27}$ & 0.13$_{-0.03}^{+0.04}$ & \\
110715A & ism & 52.35$_{-0.02}^{+0.02}$ & 2.14$_{-0.01}^{+0.01}$ & -0.99$_{-0.09}^{+0.09}$ & -0.78$_{-0.13}^{+0.11}$ & ----- & 0.18$_{-0.01}^{+0.01}$ & [27] \\
110726A & ism & 51.78$_{-0.26}^{+0.32}$ & 2.04$_{-0.02}^{+0.05}$ & -1.60$_{-0.91}^{+0.77}$ & -0.37$_{-1.16}^{+1.36}$ & ----- & 0.36$_{-0.12}^{+0.10}$ & \\
110918A & wind & 53.62$_{-0.26}^{+0.09}$ & 2.08$_{-0.07}^{+0.04}$ & -3.53$_{-1.06}^{+2.38}$ & ----- & 0.66$_{-0.77}^{+0.56}$ & 0.29$_{-0.07}^{+0.12}$ & \\
----- & ism & 53.61$_{-0.12}^{+0.13}$ & 2.05$_{-0.01}^{+0.02}$ & -3.39$_{-0.92}^{+1.12}$ & -0.29$_{-1.71}^{+1.27}$ & ----- & 0.11$_{-0.04}^{+0.05}$ & \\
111209A & ism & 53.10$_{-0.04}^{+0.04}$ & 2.02$_{-0.002}^{+0.002}$ & -0.80$_{-0.09}^{+0.09}$ & -3.66$_{-0.15}^{+0.14}$ & ----- & 0.08$_{-0.004}^{+0.004}$ & [28] \\
120327A & wind & 52.79$_{-0.08}^{+0.07}$ & 2.39$_{-0.08}^{+0.08}$ & -1.90$_{-0.51}^{+0.38}$ & ----- & -1.53$_{-0.18}^{+0.27}$ & 0.28$_{-0.13}^{+0.15}$ & \\
120404A & wind & 52.77$_{-0.23}^{+0.25}$ & 2.70$_{-0.30}^{+0.17}$ & -4.58$_{-0.33}^{+1.48}$ & ----- & -0.28$_{-0.29}^{+0.30}$ & 0.21$_{-0.05}^{+0.17}$ & [29] \\
----- & ism & 52.76$_{-1.01}^{+0.36}$ & 2.73$_{-0.26}^{+0.12}$ & -4.44$_{-0.42}^{+2.78}$ & 0.78$_{-0.79}^{+0.65}$ & ----- & 0.08$_{-0.02}^{+0.03}$ & [29] \\
120521C & ism & 53.92$_{-0.29}^{+0.35}$ & 2.08$_{-0.06}^{+0.04}$ & -4.54$_{-0.37}^{+2.80}$ & -1.36$_{-1.86}^{+0.67}$ & ----- & 0.29$_{-0.14}^{+0.14}$ & [30] \\
120729A & wind & 52.43$_{-0.71}^{+1.14}$ & 2.42$_{-0.24}^{+0.29}$ & -3.16$_{-1.29}^{+1.54}$ & ----- & -1.78$_{-1.51}^{+1.57}$ & 0.15$_{-0.08}^{+0.18}$ & \\
----- & ism & 51.79$_{-0.40}^{+0.53}$ & 2.51$_{-0.31}^{+0.19}$ & -3.57$_{-0.89}^{+1.05}$ & 0.50$_{-1.63}^{+1.06}$ & ----- & 0.08$_{-0.02}^{+0.05}$ & \\
120923A & wind & 53.35$_{-1.44}^{+1.56}$ & 2.74$_{-0.67}^{+0.19}$ & -2.66$_{-1.50}^{+1.55}$ & ----- & -3.05$_{-0.28}^{+2.56}$ & 0.29$_{-0.15}^{+0.14}$ & \\
121024A & ism & 52.82$_{-0.18}^{+0.17}$ & 2.13$_{-0.05}^{+0.28}$ & -3.25$_{-0.69}^{+1.04}$ & 0.60$_{-1.70}^{+1.04}$ & ----- & 0.10$_{-0.04}^{+0.04}$ & \\
130215A & wind & 52.57$_{-0.03}^{+1.16}$ & 2.01$_{-0.001}^{+0.002}$ & -1.76$_{-0.44}^{+0.50}$ & ----- & 0.02$_{-0.38}^{+1.88}$ & 0.45$_{-0.16}^{+0.04}$ & \\
130606A & wind & 53.73$_{-1.51}^{+0.23}$ & 2.70$_{-0.04}^{+0.05}$ & -4.64$_{-0.26}^{+2.14}$ & ----- & -0.27$_{-0.06}^{+1.87}$ & 0.32$_{-0.13}^{+0.17}$ & [31,32] \\
130831A & wind & 53.71$_{-1.00}^{+1.30}$ & 2.46$_{-0.42}^{+0.02}$ & -3.44$_{-1.07}^{+1.42}$ & ----- & -2.61$_{-0.06}^{+3.14}$ & 0.27$_{-0.11}^{+0.15}$ & \\
131030A & ism & 52.84$_{-0.05}^{+0.12}$ & 2.34$_{-0.03}^{+0.01}$ & -1.38$_{-2.41}^{+0.07}$ & -3.65$_{-0.06}^{+3.85}$ & ----- & 0.05$_{-0.002}^{+0.114}$ & \\
140311A & wind & 53.23$_{-0.10}^{+0.09}$ & 2.15$_{-0.03}^{+0.04}$ & -1.28$_{-0.46}^{+0.47}$ & ----- & -0.43$_{-0.43}^{+0.40}$ & 0.34$_{-0.13}^{+0.11}$ & [33] \\
140419A & wind & 54.33$_{-0.80}^{+1.12}$ & 2.32$_{-0.26}^{+0.04}$ & -1.12$_{-1.16}^{+0.53}$ & ----- & -3.34$_{-0.06}^{+4.14}$ & 0.31$_{-0.18}^{+0.13}$ & [34] \\
140423A & ism & 54.00$_{-0.16}^{+0.15}$ & 2.10$_{-0.02}^{+0.02}$ & -4.73$_{-0.19}^{+0.35}$ & -0.52$_{-0.30}^{+0.42}$ & ----- & 0.06$_{-0.01}^{+0.01}$ & \\
140430A & ism & 52.77$_{-0.42}^{+0.95}$ & 2.45$_{-0.11}^{+0.03}$ & -2.96$_{-1.39}^{+1.47}$ & -2.75$_{-1.58}^{+2.55}$ & ----- & 0.28$_{-0.15}^{+0.15}$ & \\
140506A & ism & 53.83$_{-0.17}^{+0.33}$ & 2.00$_{-0.002}^{+0.002}$ & -4.25$_{-0.24}^{+0.49}$ & 1.41$_{-0.95}^{+0.44}$ & ----- & 0.30$_{-0.09}^{+0.06}$ & \\
140515A & ism & 53.65$_{-0.18}^{+0.32}$ & 2.08$_{-0.05}^{+0.15}$ & -3.43$_{-0.93}^{+1.31}$ & -0.72$_{-1.95}^{+1.10}$ & ----- & 0.28$_{-0.15}^{+0.15}$ & [35] \\
140606B & ism & 51.51$_{-0.10}^{+0.19}$ & 2.11$_{-0.06}^{+0.05}$ & -2.91$_{-0.91}^{+1.44}$ & 0.11$_{-2.10}^{+1.40}$ & ----- & 0.24$_{-0.11}^{+0.11}$ & \\
140629A & ism & 52.95$_{-0.19}^{+0.24}$ & 2.25$_{-0.03}^{+0.03}$ & -4.45$_{-0.40}^{+0.63}$ & 0.24$_{-0.60}^{+0.77}$ & ----- & 0.07$_{-0.01}^{+0.02}$ & \\
140801A & ism & 52.24$_{-0.15}^{+0.20}$ & 2.18$_{-0.02}^{+0.02}$ & -3.39$_{-0.97}^{+1.65}$ & -0.16$_{-2.23}^{+1.55}$ & ----- & 0.23$_{-0.11}^{+0.13}$ & \\
150910A & wind & 53.69$_{-0.83}^{+1.60}$ & 2.42$_{-0.25}^{+0.03}$ & -2.41$_{-1.30}^{+1.50}$ & ----- & -2.85$_{-0.08}^{+1.56}$ & 0.29$_{-0.15}^{+0.13}$ & \\
160625B & ism & 54.23$_{-0.12}^{+0.11}$ & 2.27$_{-0.001}^{+0.001}$ & -4.93$_{-0.05}^{+0.10}$ & -0.90$_{-0.24}^{+0.28}$ & ----- & 0.17$_{-0.02}^{+0.02}$ & [36,37] \\
161023A & ism & 52.81$_{-0.05}^{+0.07}$ & 2.93$_{-0.01}^{+0.01}$ & -2.11$_{-0.08}^{+0.07}$ & -2.01$_{-0.07}^{+0.07}$ & ----- & 0.08$_{-0.003}^{+0.003}$ & \\
180325A & wind & 54.22$_{-0.35}^{+0.73}$ & 2.01$_{-0.001}^{+0.005}$ & -2.73$_{-0.08}^{+0.17}$ & ----- & 1.96$_{-0.06}^{+0.03}$ & 0.18$_{-0.07}^{+0.07}$ & \\
----- & ism & 52.91$_{-0.13}^{+0.33}$ & 2.60$_{-0.19}^{+0.09}$ & -3.95$_{-0.68}^{+0.94}$ & 0.37$_{-1.08}^{+1.05}$ & ----- & 0.08$_{-0.02}^{+0.03}$ & \\
180720B & wind & 54.78$_{-0.24}^{+0.25}$ & 2.18$_{-0.01}^{+0.01}$ & -4.81$_{-0.14}^{+0.37}$ & ----- & -0.59$_{-0.21}^{+0.18}$ & 0.33$_{-0.14}^{+0.12}$ & [38] \\
\hline\hline
\caption{[1] - \citet{Berger05}; [2] - \url{http://www.aoc.nrao.edu/~dfrail/allgrb_table.shtml}; [3] - \citet{Cameron05}; [4] - \citet{Cameron05b}; [5] - \citet{Cameron05c}; [6] - \citet{Soderberg06}; [7] - \citet{Chandra06}; [8] - \citet{Chandra08}; [9] - \citet{Chandra07}; [10] - \citet{vanderHorst08}; [11] - \citet{Soderberg08}; [12] - \citet{vanderHorst08b}; [13] - \citet{Kharinov10}; [14] - \citet{Guidorzi11}; [15] - \citet{Soderberg08b}; [16] - \citet{Melandri10}; [17] - \citet{vanderHorst09}; [18] - \citet{Cenko11}; [19] - \citet{Chandra10}; [20] - \citet{Frail09}; [21] - \citet{Moin13}; [22] - \citet{Pooley10}; [23] - \citet{Chandra12}; [24] - \citet{Frail11}; [25] - \citet{vanderHorst11}; [26] - \citet{Zauderer11}; [27] - \citet{Hancock11}; [28] - \citet{Hancock12}; [29] - \citet{Zauderer12}; [30] - \citet{Laskar14}; [31] - \citet{Laskar13}; [32] - \citet{Castro-Tirado13}; [33] - \citet{Laskar14b}; [34] - \citet{Perley14}; [35] - \citet{Laskar14c}; [36] - \citet{Alexander16}; [37] - \citet{Mooley16}; [38] - \citet{Chandra18}}
\end{longtable}
\end{center}

\begin{figure}
    \centering
    \includegraphics[width=14cm]{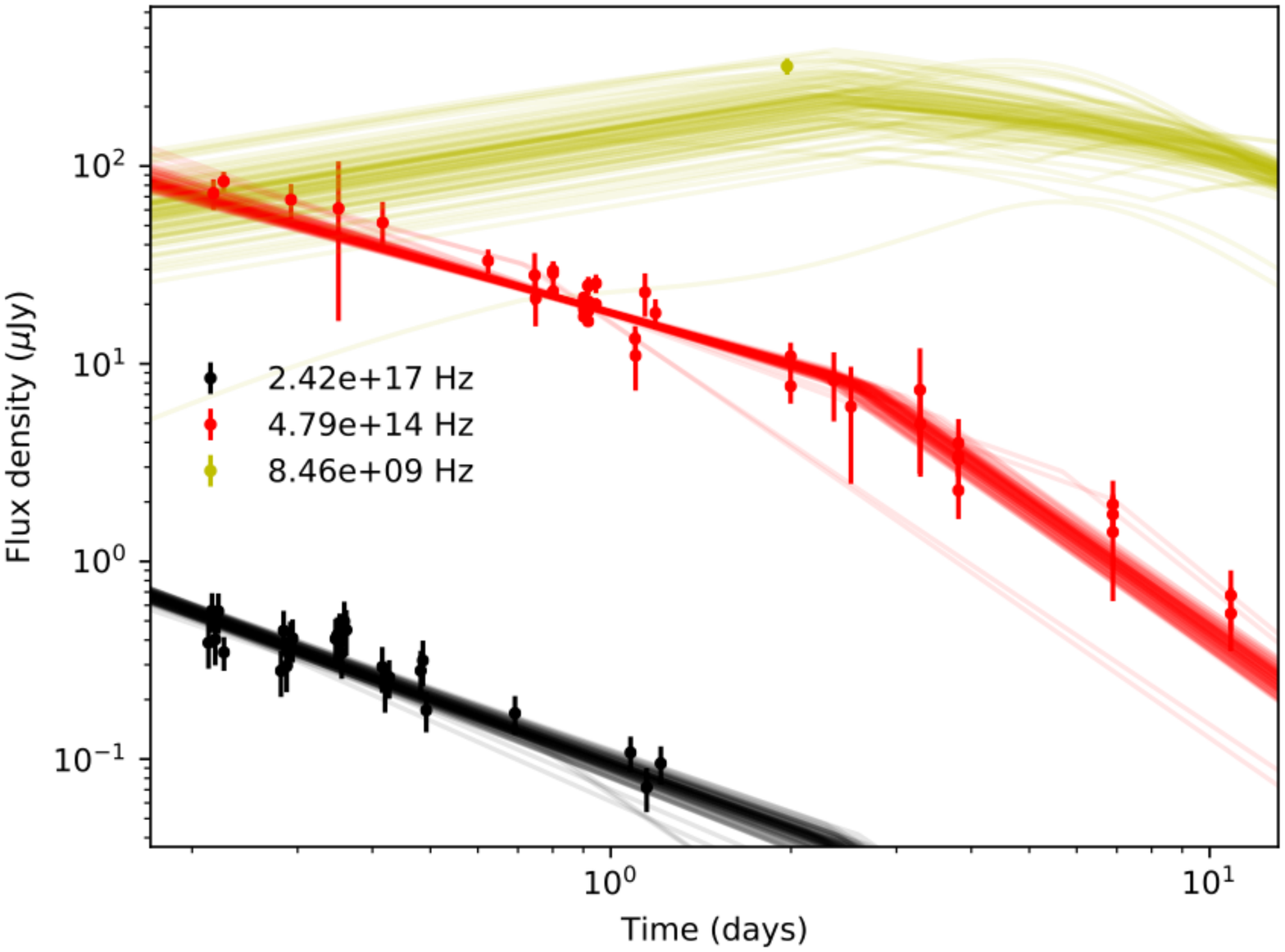}
    \includegraphics[width=14cm]{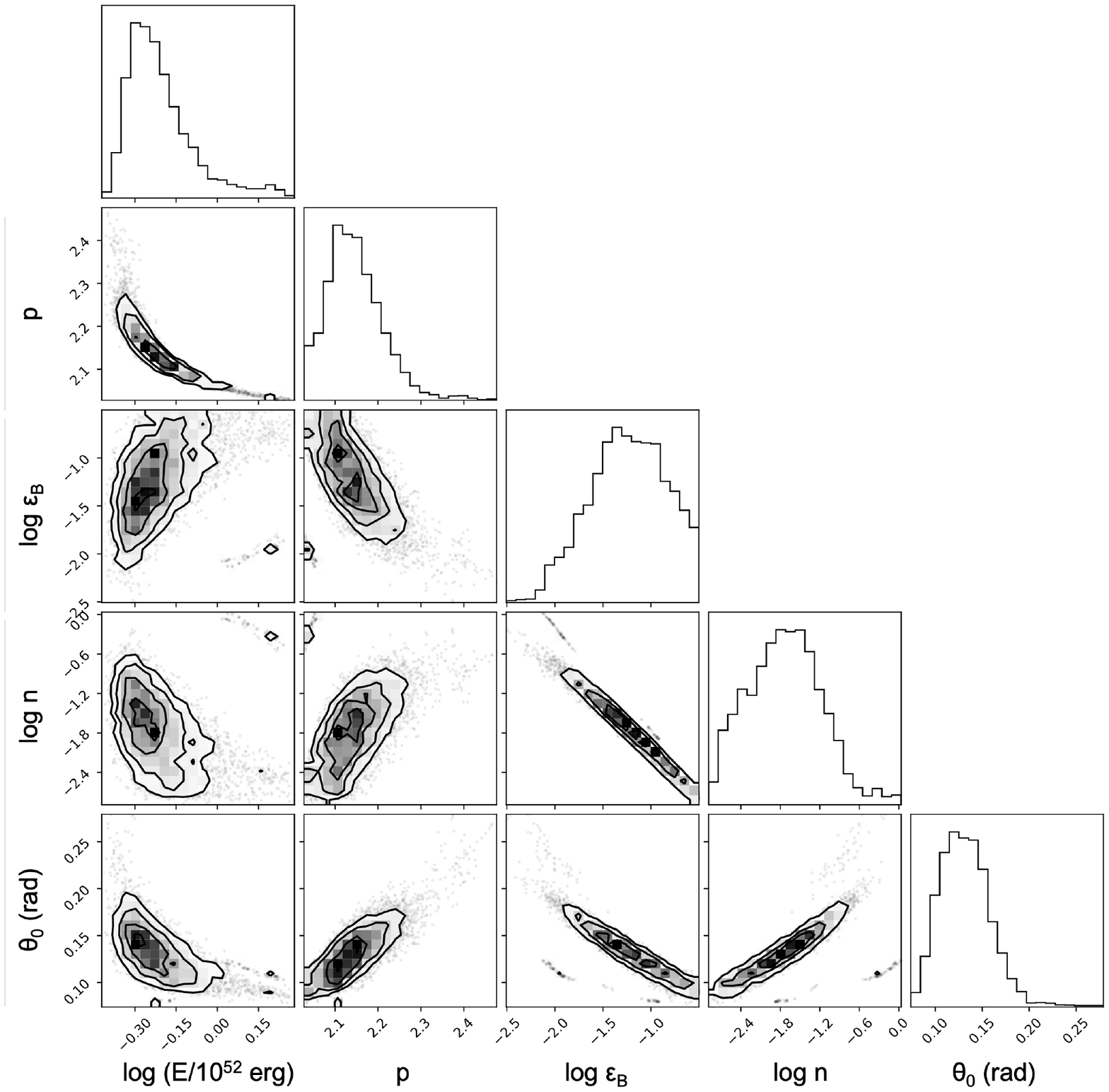}
    \caption{Light curve and corner plot for the fit to GRB 081007 (ISM).}
    \label{app:081007}
\end{figure}

\begin{figure}
    \centering
    \includegraphics[width=14cm]{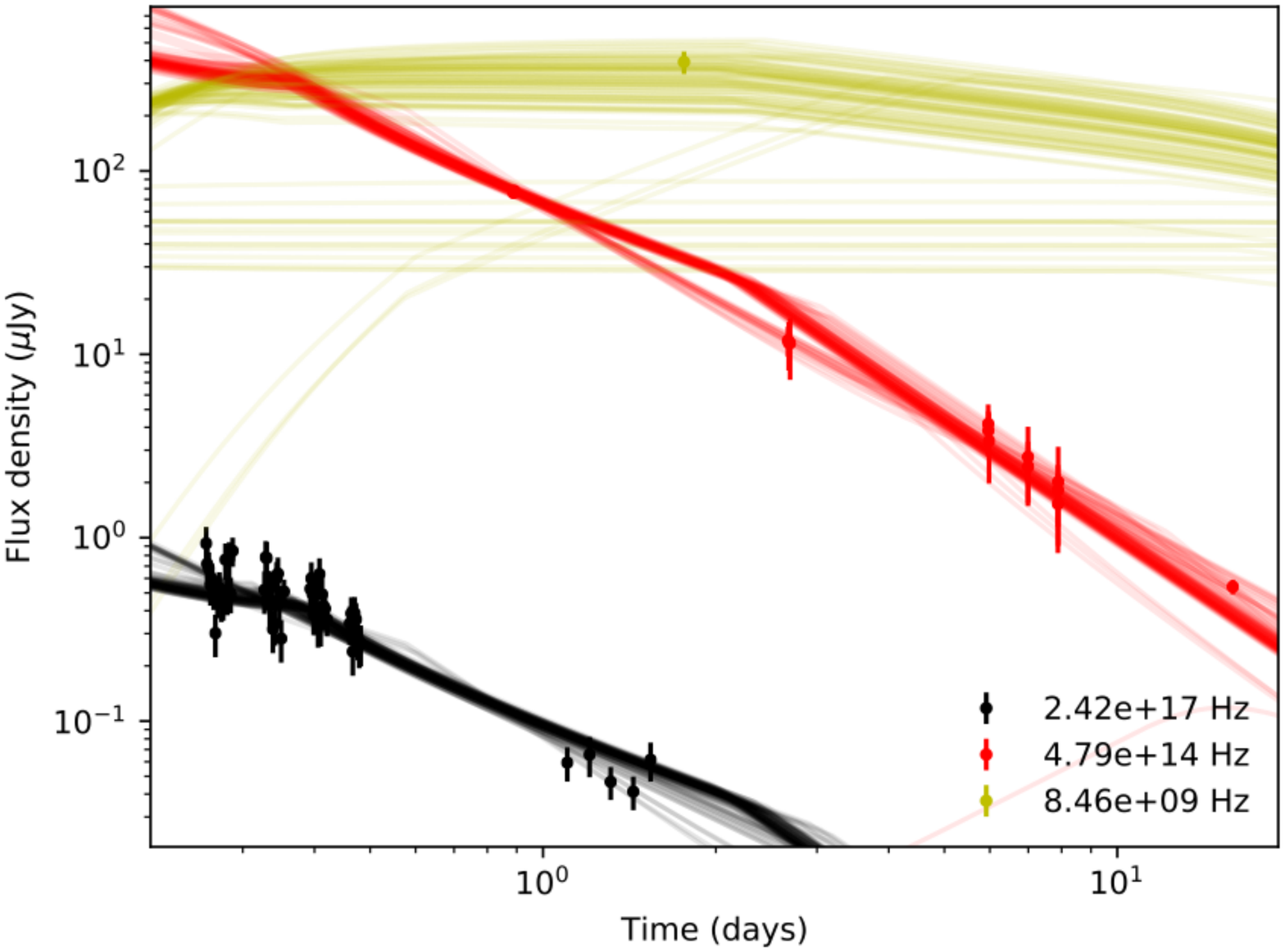}
    \includegraphics[width=14cm]{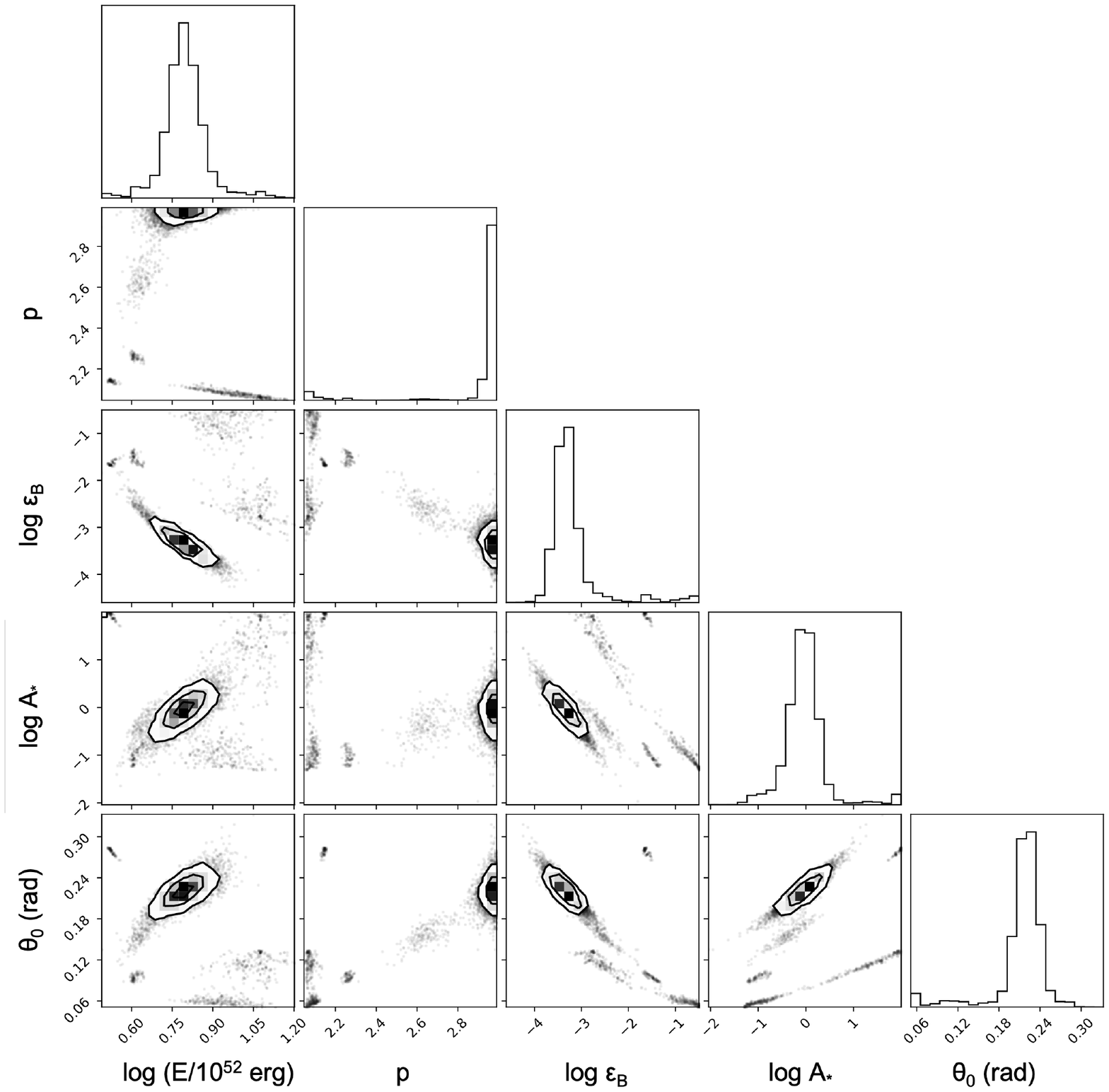}
    \caption{Light curve and corner plot for the fit to GRB 071003 (wind).}
    \label{app:071003}
\end{figure}


\bsp	
\label{lastpage}
\end{document}